\documentstyle[12pt]{article}

\def\hybrid{\topmargin -25pt    \oddsidemargin 0pt
        \headheight 0pt \headsep 0pt
        \textwidth 6.25in       
        \textheight 9.5in       
        \marginparwidth .875in
        \parskip 5pt plus 1pt   \jot = 1.5ex}
\def\cQ{{\cal Q}}
\def\cG{{\cal G}}
\def\cL{{\cal L}}
\def\cH{{\cal H}}
\def\ket#1{|{#1}\rangle}
\def\noi{\noindent}
\def\half{{1\over2}}
\def\baselinestretch{1.2}

\catcode`\@=11

\def\marginnote#1{}
\def\draftlabel#1{{\@bsphack\if@filesw {\let\thepage\relax
   \xdef\@gtempa{\write\@auxout{\string
      \newlabel{#1}{{\@currentlabel}{\thepage}}}}}\@gtempa
   \if@nobreak \ifvmode\nobreak\fi\fi\fi\@esphack}
        \gdef\@eqnlabel{#1}}
\def\@eqnlabel{}
\def\@vacuum{}
\def\draftmarginnote#1{\marginpar{\raggedright\scriptsize\tt#1}}

\def\draft{\oddsidemargin -.2truein
        \def\@oddfoot{\sl preliminary draft \hfil
        \rm\thepage\hfil\sl\today\quad\militarytime}
        \let\@evenfoot\@oddfoot \overfullrule 3pt
        \let\label=\draftlabel
        \let\marginnote=\draftmarginnote
   \def\@eqnnum{(\theequation)\rlap{\kern\marginparsep\tt\@eqnlabel}%
\global\let\@eqnlabel\@vacuum}  }

\def\preprint{\twocolumn\sloppy\flushbottom\parindent 2em
        \leftmargini 2em\leftmarginv .5em\leftmarginvi .5em
        \oddsidemargin -.5in    \evensidemargin -.5in
        \columnsep .4in \footheight 0pt
        \textwidth 10.in        \topmargin  -.6in
        \headheight 12pt \topskip .3in
        \textheight 6.9in \footskip 0pt
        \def\@oddhead{\thepage\hfil\addtocounter{page}{1}\thepage}
        \let\@evenhead\@oddhead \def\@oddfoot{} \def\@evenfoot{} }

\def\numberbysection{\@addtoreset{equation}{section}
        \def\theequation{\thesection.\arabic{equation}}}

\def\underline#1{\relax\ifmmode\@@underline#1\else
        $\@@underline{\hbox{#1}}$\relax\fi}

\def\titlepage{\@restonecolfalse\if@twocolumn\@restonecoltrue
\onecolumn
     \else \newpage \fi \thispagestyle{empty}\c@page\z@
        \def\thefootnote{\fnsymbol{footnote}} }

\def\endtitlepage{\if@restonecol\twocolumn \else \newpage \fi
        \def\thefootnote{\arabic{footnote}}
        \setcounter{footnote}{0}}  

\catcode`@=12
\relax

\def\figcap{\section*{Figure Captions\markboth
        {FIGURECAPTIONS}{FIGURECAPTIONS}}\list
        {Figure \arabic{enumi}:\hfill}{\settowidth\labelwidth{Figure
999:}
        \leftmargin\labelwidth
        \advance\leftmargin\labelsep\usecounter{enumi}}}
 \relax
\def\tablecap{\section*{Table Captions\markboth
        {TABLECAPTIONS}{TABLECAPTIONS}}\list
        {Table \arabic{enumi}:\hfill}{\settowidth\labelwidth{Table
999:}
        \leftmargin\labelwidth
        \advance\leftmargin\labelsep\usecounter{enumi}}}
 \relax
\def\reflist{\section*{References\markboth
        {REFLIST}{REFLIST}}\list
        {[\arabic{enumi}]\hfill}{\settowidth\labelwidth{[999]}
        \leftmargin\labelwidth
        \advance\leftmargin\labelsep\usecounter{enumi}}}
 \relax
\makeatletter
\newcounter{pubctr}
\def\publist{\@ifnextchar[{\@publist}{\@@publist}}
\def\@publist[#1]{\list
        {[\arabic{pubctr}]\hfill}{\settowidth\labelwidth{[999]}
        \leftmargin\labelwidth
        \advance\leftmargin\labelsep
        \@nmbrlisttrue\def\@listctr{pubctr}
        \setcounter{pubctr}{#1}\addtocounter{pubctr}{-1}}}
\def\@@publist{\list
        {[\arabic{pubctr}]\hfill}{\settowidth\labelwidth{[999]}
        \leftmargin\labelwidth
        \advance\leftmargin\labelsep
        \@nmbrlisttrue\def\@listctr{pubctr}}}
 \relax
\makeatother
\newskip\humongous \humongous=0pt plus 1000pt minus 1000pt

\newif\ifdtup

\font\Scbig=cmss10 scaled\magstep1
\font\Scscr=cmss8 scaled\magstep1
\font\Scscrscr=cmss8
\newfam\Scfam
\textfont\Scfam=\Scbig
\scriptfont\Scfam=\Scscr
\scriptscriptfont\Scfam=\Scscrscr
\def\Sc{\fam\Scfam}

\relax
\hybrid
\def\lvm{\leavevmode\hbox to\parindent{\hfill}}

\def\thefootnote{\fnsymbol{footnote}}
\def\BE{\begin{equation}}
\def\EE{\end{equation}}
\def\BA{\begin{eqnarray}}
\def\EA{\end{eqnarray}}

\def\D{\Delta}

\def\a{\alpha}
\def\th{\theta}

\def\P{\Phi}

\def\e{\epsilon}

\def\tt{\bar\tau}

\def\lvm{\leavevmode\hbox to\parindent{\hfill}}
\def\bar{\overline}
\def\req#1{(\ref{#1})}

\def\L{\left}
\def\R{\right}

\def\BE{\begin{equation}}
\def\EE{\end{equation} \vskip 0.30\baselineskip}
\def\BA{\begin{array}}
\def\EA{\end{array}}

\def\noi{\noindent}

\def\frac#1#2{{\textstyle{{#1}\over{#2}}}}
\def\half{{1\over2}}

\def\Kr#1{\delta_{{#1},0}}
\def\ket#1{|{#1}\rangle}

\def\ccases#1#2{\L\{\!\new\BA{l}{#1}\\ {#2}\EA\R.}

\def\cA{{\cal A}}
\def\cG{{\cal G}}
\def\cH{{\cal H}}

\def\cL{{\cal L}}

\def\cQ{{\cal Q}}

\def\cM{{\cal M}}
\def\cU{{\cal U}}

\def\open#1{\mbox{{\bf{#1}}}}

\def\oZ{{\open Z}}

\def\ctop{{\Sc c}}
\def\htop{{\Sc h}}

\def\a{\alpha}
\def\b{\beta}
\def\g{\gamma}
\def\Ups{\Upsilon}

\def\ie{{\it i.e.}}

\def\Qz{\cQ_0}
\def\Gz{\cG_0}
\def\Qn{$\Qz$}
\def\Gn{$\Gz$}
\def\kc{{\ket{\chi}}}

\newif\ifold \oldtrue \def\new{\oldfalse}
\let\ssection=\section
\def\section{\setcounter{equation}{0}\ssection}

\begin{document}
\renewcommand{\theequation}{\thesection.\arabic{equation}}
\newcommand{\beq}{\begin{equation}}
\newcommand{\eeq}[1]{\label{#1}\end{equation}}
\newcommand{\ber}{\begin{eqnarray}}
\newcommand{\eer}[1]{\label{#1}\end{eqnarray}}
\begin{titlepage}
\begin{center}
\hfill IMAFF-FM-97/02\\
\hfill NIKHEF-97-022\\
\hfill hep-th/9706041
\vskip .4in

{\large \bf Chiral Determinant Formulae and Subsingular Vectors for  
 the N=2 Superconformal Algebras}
\vskip .4in

{\bf Beatriz Gato-Rivera}$^{a,b}$ {\bf and Jose Ignacio Rosado}$^a$\\
\vskip .3in

${\ }^a$ 
{\em Instituto de Matem\'aticas y
 F\'\i sica Fundamental, CSIC,\\ Serrano 123,
Madrid 28006, Spain} \footnote{e-mail address:
bgato@pinar1.csic.es}\\
\vskip .25in

${\ }^b$
 {\em NIKHEF, Kruislaan 409, NL-1098 SJ Amsterdam, The Netherlands}\\

\end{center}

\vskip .4in

\begin{center} {\bf ABSTRACT } \end{center}
\begin{quotation}

We derive conjectures for the N=2 ``chiral" determinant formulae of the 
Topological algebra, the Antiperiodic NS algebra, and the Periodic R algebra,  
corresponding to incomplete Verma modules built on chiral topological 
primaries, chiral and antichiral NS primaries, and Ramond ground states, 
respectively. Our method is based on the analysis of the singular vectors 
in chiral Verma modules and their spectral flow symmetries, together with 
some computer exploration and some consistency checks. In addition, and as a 
consequence, we uncover the existence of subsingular vectors in these 
algebras, giving examples (subsingular vectors are non-highest-weight null 
vectors which are not descendants of any highest-weight singular vectors).    

\end{quotation}
\vskip .15in

October 1997\\  
 
\end{titlepage}
\vfill
\eject
\def\baselinestretch{1.2}
\baselineskip 17 pt
\section{Introduction and Notation}\lvm

The N=2 Superconformal algebras provide the symmetries underlying the
N=2 strings \cite{Ade} \cite{FrTs} \cite{OoVa} \cite{Marcus}. 
In addition, the topological 
version of the algebra is realized in the world-sheet of the bosonic
string \cite{BeSe2}, as well as in the world-sheet of the  
superstrings \cite{BLNW}.
Various aspects concerning singular vectors of the 
N=2 Superconformal algebras in chiral Verma modules 
have been studied in several papers during
the last few years, most of them involving the Topological 
algebra. For example, in \cite{BeSe2} and \cite{BeSe3} it was shown
that the uncharged BRST-invariant singular vectors, in the ``mirror
bosonic string" realization (KM) of the Topological algebra, are
related to the Virasoro constraints on the KP $\tau$-function. In
\cite{Sem} an isomorphism was uncovered between the uncharged
BRST-invariant singular vectors and the singular vectors of the
sl(2) Kac-Moody algebra. Some properties of the topological singular 
vectors in the DDK and KM realizations were analyzed in \cite{BJI2}. 
In \cite{BJI3} the complete set of topological singular vectors at 
level 2 (four types) was written down, together with the spectral
flow automorphism $\cA$ which transforms all kinds of topological singular
vectors into each other. In \cite{BJI6} the family structure of  
the topological singular vectors was analyzed (not only in chiral
Verma modules but also in complete Verma modules, where there are
thirty-three different types of topological singular vectors).
 
 In all of those papers
some explicit examples of singular vectors in chiral
Verma modules were written down,
ranging from level 1 until level 4. In addition, in \cite{Sem}
a general formula was given for the spectrum of U(1) charges
corresponding to the topological chiral Verma modules which contain 
uncharged BRST-invariant singular vectors. Although the formula fitted
with the known data, a proof or derivation was lacking\footnote{In 
addition that formula, eq. (3.1) in ref. \cite{Sem},
was presented not as a conjecture but as a
straightforward derivation from the determinant formula of the N=2
Antiperiodic NS algebra, in particular from the uncharged series 
in ref. \cite{BFK}, which is not the case since the
determinant formulae does not apply to 
incomplete Verma modules with constraints. 
The fact that the formula given in ref. \cite{Sem} cannot be derived
from the determinant formula in ref. \cite{BFK}, as claimed by the
author, is precisely what leads to the discovery of subsingular
vectors, as we will see.}. 
Similarly, general formulae for the spectrum of conformal weights
and/or U(1) charges for chiral Verma modules of the Antiperiodic NS
algebra which contain singular vectors are absent, and the same is 
true for the Verma modules of the Periodic R algebra built on the Ramond
ground states. 
These spectra, which cannot be obtained directly from the 
roots of the N=2 determinant formulae \cite{BFK} \cite{Nam} \cite{KaMa3}, 
since these only apply to complete Verma modules without
constraints, can be viewed rather as the roots of the ``chiral"
determinant formulae of the N=2 Superconformal algebras (the Ramond 
ground states are directly related to the chiral NS primary states via 
the spectral flows).  

In this paper we present conjectures for the N=2 ``chiral" determinant 
formulae corresponding to the chiral Verma modules of the 
Topological algebra, the chiral Verma modules of the Antiperiodic NS 
algebra, and the Verma modules of the Periodic R algebra built on 
the Ramond ground states. 
In the absence of rigorous proofs, we have checked 
our results from level 1/2 to level 4 and, in addition, we provide some  
consistency checks. 

We have proceeded in two steps. First we have derived conjectures
for the roots of the N=2 chiral determinants, using some properties
of the singular vectors in chiral Verma modules, together with the ansatz
that the roots of the N=2 chiral determinants are contained in the set
of roots of the N=2 determinants, and in the simplest possible way,
in addition. 
Second, we have written down the N=2 chiral determinant formulae
using the conjectures for the roots, some computer exploration, and
the consistency checks.
Our conjectures for the roots of the N=2 chiral
determinants imply that there exist subsingular vectors, \ie\
singular vectors in the chiral Verma modules which are not singular
in the complete Verma modules, where they are non-highest weight null 
vectors not descendant of any singular vectors. To fully understand this 
issue \cite{FF} one has to take into account that the chiral Verma 
modules are nothing but the quotient of complete Verma modules by 
submodules generated by singular vectors.

This paper is organized as follows.
In section 2 we first explain the family structure of the singular
vectors of the Topological algebra built on 
chiral topological primaries. Then
we derive the direct relation between these topological 
singular vectors and
the singular vectors of the Antiperiodic NS algebra built on
chiral primaries. This relation implies, using the family structure 
of the topological singular vectors,
that the charged and uncharged NS singular vectors built
on chiral primaries must come in pairs, although in different
Verma modules related to each other by the spectral flow
automorphism of the Topological algebra \cite{BJI3}.

We use this result in section 3 to derive  
conjectures for the spectrum of U(1) charges $\htop$
(and conformal weights $\Delta$) corresponding to the chiral  
Verma modules (NS and topological) which contain singular vectors.
These spectra are the roots of the N=2 chiral determinant formulae. 
To derive the conjectures we also make the ansatz 
that the roots of the N=2 chiral determinants coincide with the 
roots of the N=2 determinants (\ie\ for complete Verma modules) 
specialized to the values of (or relations between) the conformal
weight $\Delta$ and U(1) charge $\htop$ which occur in chiral
Verma modules. Namely, $\Delta=0$ for the case of the Topological
algebra, and $\Delta=\pm{\htop\over2}$ for the case of 
the Antiperiodic NS algebra (+ for chiral representations and
$-$ for antichiral representations).  
Actually we only need to work out the ansatz
for one of the algebras since the ansatz for the other algebra 
follows automatically via the relations between the corresponding
singular vectors. 
We have chosen the Antiperiodic NS algebra for
this purpose. The fact that charged and uncharged singular vectors come
in pairs then implies that
half of the zeroes of the quadratic vanishing surface
$f^A_{r,s}=0$, for every pair $(r,s)$,
correspond to uncharged singular vectors at level $rs\over2$,
and the other half correspond to charged singular vectors
at level $r(s+2)-1\over2$
(in addition to the zeroes of the vanishing plane $g_k^A = 0$).
This contrasts sharply with the spectra of singular
vectors for complete Verma modules,
for which all the zeroes of the quadratic vanishing surface 
$f_{r,s}^A=0$ correspond to uncharged singular vectors at level 
$rs\over2$, and implies that the ``new" charged singular vectors
which appear in the chiral Verma modules are subsingular vectors.

In section 4 we derive the analogous conjecture for the Periodic 
R algebra, that is the spectrum of U(1) charges $\htop$ corresponding 
to the R Verma modules, built on the Ramond ground states (for which
$\D={\ctop\over24}$), which contain singular vectors. For this we only
need to use the spectral flows, with the appropriate parameters which
map singular vectors of the Antiperiodic
NS algebra, built on chiral primaries, into singular vectors of the
Periodic R algebra, built on the Ramond ground states.

In section 5 we finally write down expressions for the N=2 chiral 
determinant formulae, using the conjectures for the roots together
with some computer investigation and some consistency checks.
Section 6 is devoted to conclusions and final remarks.

Finally, in the Appendix we analyze thoroughly 
 the NS singular vectors at levels 1 and $3\over2$. 
 We write down all the equations 
resulting from the highest weight conditions, showing
that these equations, and therefore their solutions, 
are different when imposing or not chirality on the
primary states on which the singular vectors are built.
Namely, when the primary state is non-chiral all the 
solutions of the quadratic vanishing surface $f_{1,2}^A=0$
correspond to level 1 uncharged singular vectors. When the
primary state is chiral, however, half of the solutions of $f_{1,2}^A=0$
specialized to the cases $\Delta = \pm {\htop\over2}$
 correspond to level $3\over2$ charged singular vectors. 
We show that these charged singular vectors become
non-highest weight null vectors,
once the chirality on the primary state is switched off,
which are not descendants of any h.w. singular vector, \ie\
they are subsingular vectors. 

\vskip .12in
\subsection*{Notation}

\vskip .15in
\noi
{\it Primary states} denote non-singular highest weight (h.w.) vectors.

\noi
{\it Null vectors} are zero-norm states (not necessarily h.w.).

\noi
{\it Singular vectors} are h.w. null vectors, equivalently h.w. descendant 
states.

\noi
{\it Subsingular vectors} are non-h.w. null vectors not descendants of
any singular vectors, they become singular (\ie\ h.w.) in the quotient 
of the Verma module by a submodule generated by singular vectors.

\noi
The Antiperiodic NS algebra will be denoted as the {\it NS algebra}. The
chiral and antichiral primary states and Verma modules
will be denoted simply as {\it chiral}, unless otherwise indicated.

\noi
The Periodic R algebra will be denoted as the {\it R algebra}.

\noi
The singular vectors of the Topological algebra
will be denoted as {\it topological singular vectors} $\ket{\chi_T}$.

\noi
The singular vectors of the NS algebra
will be denoted as {\it NS singular vectors} $\ket{\chi_{NS}}$.

\noi
The singular vectors of the R algebra
will be denoted as {\it R singular vectors} $\ket{\chi_R}$.

\vskip .55in
\section{Some Properties of Singular Vectors in Chiral Verma Modules}\lvm

\subsection{Singular Vectors of the Topological Algebra in Chiral
Verma Modules}\lvm

{\it The Topological Algebra}.$-$ The algebra obtained by applying the
topological twists \cite{EY}, \cite{W-top} on 
the N=2 Superconformal algebra reads  \cite{DVV}

\BE\new\BA{lclclcl}
\L[\cL_m,\cL_n\R]&=&(m-n)\cL_{m+n}\,,&\qquad&[\cH_m,\cH_n]&=
&{\ctop\over3}m\Kr{m+n}\,,\\
\L[\cL_m,\cG_n\R]&=&(m-n)\cG_{m+n}\,,&\qquad&[\cH_m,\cG_n]&=&\cG_{m+n}\,,
\\
\L[\cL_m,\cQ_n\R]&=&-n\cQ_{m+n}\,,&\qquad&[\cH_m,\cQ_n]&=&-\cQ_{m+n}\,,\\
\L[\cL_m,\cH_n\R]&=&\multicolumn{5}{l}{-n\cH_{m+n}+{\ctop\over6}(m^2+m)
\Kr{m+n}\,,}\\
\L\{\cG_m,\cQ_n\R\}&=&\multicolumn{5}{l}{2\cL_{m+n}-2n\cH_{m+n}+
{\ctop\over3}(m^2+m)\Kr{m+n}\,,}\EA\qquad m,~n\in\oZ\,.\label{topalgebra}
\EE

\noi
where $\cL_m$ and $\cH_m$ are the bosonic generators corresponding
to the energy momentum tensor (Virasoro generators)
 and the topological $U(1)$ current respectively, while
$\cQ_m$ and $\cG_m$ are the fermionic generators corresponding
to the BRST current and the spin-2 fermionic current
 respectively. The ``topological central
charge" $\ctop$ is the central charge of the untwisted N=2
Superconformal algebra. This algebra is topological because the
Virasoro generators can be expressed as $\cL_m=\half \{\cG_m,\Qz \}$,
where \Qn\ is the BRST charge. This implies, as is well known, that
the correlators of fields do not depend on the metric.

\vskip .17in

{\it The Topological Twists}.$-$ The Topological algebra is satisfied 
by the two sets of topological generators

\BE\new\BA{rclcrcl}
\cL^{(1)}_m&=&\multicolumn{5}{l}{L_m+\half(m+1)H_m\,,}\\
\cH^{(1)}_m&=&H_m\,,&{}&{}&{}&{}\\
\cG^{(1)}_m&=&G_{m+\half}^+\,,&\qquad &\cQ_m^{(1)}&=&G^-_{m-\half}
\,,\label{twa}\EA\EE

\noi
and

\BE\new\BA{rclcrcl}
\cL^{(2)}_m&=&\multicolumn{5}{l}{L_m-\half(m+1)H_m\,,}\\
\cH^{(2)}_m&=&-H_m\,,&{}&{}&{}&{}\\
\cG^{(2)}_m&=&G_{m+\half}^-\,,&\qquad &
\cQ_m^{(2)}&=&G^+_{m-\half}\,,\label{twb}\EA\EE

\noi
corresponding to the two possible twistings of the superconformal generators
$L_m, H_m, G^{+}_m$ and $G^{-}_m$. 
The topological twists, which we denote as $T_{W1}$ and $T_{W2}$, are
mirrored under the interchange $H_m\leftrightarrow -H_m$,
$G_r^+\leftrightarrow G_r^-$, as one can see.

\vskip .17in

{\it Chiral Topological Primaries}.$-$
Of special importance are the topological primary states annihilated 
by both \Gn\ and \Qn, denoted as ``chiral". The 
anticommutator $\{\cG_0,\cQ_0\}=2\cL_0$ shows that these 
states have zero conformal weight $\D$, therefore their only quantum
number is 
their U(1) charge $\htop$ ($\cH_0\ket\Phi=\htop\ket\Phi$). The secondary
states built on chiral primaries have positive conformal weight $\D=l>0$,
as a result, where $l$ is the level of the state. This anticommutator
also shows that any secondary state with conformal
weight $\D\neq 0$ can be decomposed 
 into a $\cQ_0$-closed state and a $\cG_0$-closed state,  
and also implies that
a $\cQ_0$-closed secondary state with $\D\neq0$ is $\cQ_0$-exact as
well (and similarly with $\cG_0$). 
Therefore, the chiral primaries are physical states 
(BRST-closed but not BRST-exact), whereas the secondary states built 
on chiral primaries are not physical.

Regarding the twists \req{twa} and \req{twb}, a key observation is that
$(G^{+}_{1/2}, G^{-}_{-1/2})$ results in $(\cG^{(1)}_0, \cQ^{(1)}_0)$ 
and $(G^{-}_{1/2}, G^{+}_{-1/2})$ gives $(\cG^{(2)}_0, \cQ^{(2)}_0)$.
This brings about two important consequences. First, the chiral
topological primaries $\ket\P^{(1)}$ and $\ket\P^{(2)}$ correspond to
the antichiral primaries (\ie\ $G_{-1/2}^-\ket\P^{(1)}=0$) and to the
chiral primaries (\ie\ $G_{-1/2}^+\ket\P^{(2)}=0$) of the 
NS algebra, respectively. Second, one of the highest weight (h.w.)
conditions $G^{\pm}_{1/2}\ket{\chi_{NS}}=0$, of the NS algebra,
read $\cG_0\ket{\chi_T}=0$ after the corresponding twistings. Therefore,
any h.w. state (primary or secondary) of the NS algebra
results in a topological state annihilated by \Gn, under the twistings.
We will discuss in detail this issue in next subsection.

\vskip .17in

{\it Topological Secondaries}.$-$ A topological secondary, or descendant, 
state can also be labeled by its level $l$ (the conformal weight 
for states built on chiral topological primaries),
and its U(1) charge $(\htop + q)$
(the $\cH_0$-eigenvalue). It is convenient to split the total U(1) charge 
into two pieces: the U(1) charge $\htop$ of the primary state on which the 
descendant is built, thus labeling the corresponding Verma module
 $V_T(\htop)$, and the
 ``relative" U(1) charge $q$, corresponding to the 
operator acting on the primary state, given by the number of $\cG$ modes 
minus the number of $\cQ$ modes in each term. We will denote
the $\cQ_0$-closed and
$\cG_0$-closed topological secondary states
as $\ket{\chi_T}^{(q)Q}$ and 
$\ket{\chi_T}^{(q)G}$ respectively (notice that $(q)$ refers to the
relative U(1) charge of the state). Usually we will also indicate the
level $l$ and/or the Verma module $\htop$.

\vskip .17in

{\it Topological Singular Vectors on Chiral Primaries}.$-$
It turns out that the topological singular vectors built on chiral
primaries come only in four types\footnote{A rigorous proof
will be presented in a revised version of ref. \cite{BJI6}.}  
\cite{BJI6}: $\ket{\chi_T}^{(0)G}$, $\ket{\chi_T}^{(0)Q}$, 
$\ket{\chi_T}^{(1)G}$ and $\ket{\chi_T}^{(-1)Q}$. These four 
types of singular vectors can be mapped into each other by using
$\cG_0$, $\cQ_0$ and the spectral flow automorphism of the
Topological algebra, denoted by 
$\cA$. The action of $\cG_0$ and $\cQ_0$ results in
singular vectors in the same Verma module (by definition):

$$\cQ_0\ket{\chi_T}_{l,\,\htop}^{(0)G}\rightarrow
\ket{\chi_T}_{l,\,\htop}^{(-1)Q}, \qquad 
\cG_0\ket{\chi_T}_{l,\,\htop}^{(-1)Q}\rightarrow
\ket{\chi_T}_{l,\,\htop}^{(0)G}, $$
$$\cQ_0\ket{\chi_T}_{l,\,\htop}^{(1)G}
\rightarrow\ket{\chi_T}_{l,\,\htop}^{(0)Q}, \qquad 
\cG_0\ket{\chi_T}_{l,\,\htop}^{(0)Q}
\rightarrow\ket{\chi_T}_{l,\,\htop}^{(1)G}.$$

\noi
The level of the vectors
does not change under the action of $\cG_0$ and 
$\cQ_0$, obviously. Since the Verma module does not change 
either, charged and uncharged topological singular
vectors, with different BRST-invariance properties,
come always in pairs in the same Verma module.
Namely, singular vectors 
of the types $\ket{\chi_T}^{(0)Q}$ and $\ket{\chi_T}^{(1)G}$ 
are together in the same Verma module at the same level, 
and a similar statement holds for the 
singular vectors of the types $\ket{\chi_T}^{(0)G}$ and 
$\ket{\chi_T}^{(-1)Q}$.

 The spectral flow automorphism of the
 topological algebra, on the other hand, given by \cite{BJI3}

\BE\new\BA{rclcrcl}
\cA \, \cL_m \, \cA&=& \cL_m - m\cH_m\,,\\
\cA \, \cH_m \, \cA&=&-\cH_m - {\ctop\over3} \delta_{m,0}\,,\\
\cA \, \cQ_m \, \cA&=&\cG_m\,,\\
\cA \, \cG_m \, \cA&=&\cQ_m\,,\
\label{autom} \EA\EE

\noi
with $\cA^{-1} = \cA$,
changes the Verma module of the vectors as
 $V_T(\htop)\rightarrow V_T(-\htop-{\ctop\over 3}$). 
In addition, $\cA$ reverses 
the relative charge as well as the BRST-invariance properties
 of the vectors, leaving the 
level invariant. Therefore the action of $\cA$ results in the following 
mappings \cite{BJI3}

\BE
\cA\ket{\chi_T}_{l,\,\htop}^{(0)G}\rightarrow
\ket{\chi_T}_{l,-\htop-{\ctop\over3}}^{(0)Q}\,
,\qquad 
\cA\ket{\chi_T}_{l,\,\htop}^{(-1)Q}\rightarrow
\ket{\chi_T}_{l,-\htop-{\ctop\over3}}^{(1)G}\,.
\EE

\vskip .17in

{\it Family Structure}.$-$ As a consequence of the mappings given by \Gn ,  
\Qn\ and $\cA$, the topological singular vectors built on chiral primaries
come in families of four \cite{BJI6}, one of each kind at the same level.
Two of them, one charged and one uncharged, belong to
the Verma module $V_T(\htop)$, whereas the other pair belong to a
different Verma module 
$V_T(-\htop-{\ctop\over 3})$, as the diagram shows. 

\vskip .2in

\def\xgo  {\mbox{$\ket{\chi_T}_{l,\,\htop}^{(0)G} $}}
\def\xqo  {\mbox{$\ket{\chi_T}_{l,-\htop-{\ctop\over3}}^{(0)Q} $}}
\def\xqm  {\mbox{$\ket{\chi_T}_{l,\,\htop}^{(-1)Q} $}}
\def\xgp  {\mbox{$\ket{\chi_T}_{l,-\htop-{\ctop\over3}}^{(1)G} $}}

  \begin{equation}  
  \begin{array}{rcl} \xgo &
  \stackrel{\Qz}{\mbox{------}\!\!\!\longrightarrow}
  & \xqm \\[3 mm]
   \cA\,\updownarrow\ && \ \updownarrow\, \cA
  \\[3 mm]  \xqo \! & \stackrel{\Gz}
  {\mbox{------}\!\!\!\longrightarrow} & \! \xgp  \end{array}
\label{dia}  \end{equation}

\noi
The arrows \Qn\ and \Gn\ can be reversed using \Gn\ and \Qn , 
respectively (up to a constant).
For $\htop=-{\ctop\over6}$ the two Verma modules related by the
spectral flow automorphism coincide.
 Therefore, if there are singular vectors for this value of $\htop$
(see in section 3), they must come four by four in the same Verma module: 
one of each kind at the same level.

\subsection{Untwisting the Topological Singular Vectors}\lvm

The relation between the Topological algebra and the NS 
algebra is given by the topological twists
$T_{W1}$ \req{twa} and $T_{W2}$ \req{twb}.
From the purely formal point of view, the
Topological algebra \req{topalgebra} is simply a rewriting of the  NS
algebra, given by \cite{Ade}, \cite{PDiV}, \cite{LVW}, \cite{Kir1}

\BE\new\BA{lclclcl}
\L[L_m,L_n\R]&=&(m-n)L_{m+n}+{\ctop\over12}(m^3-m)\Kr{m+n}
\,,&\qquad&[H_m,H_n]&=
&{\ctop\over3}m\Kr{m+n}\,,\\
\L[L_m,G_r^\pm
\R]&=&\L({m\over2}-r\R)G_{m+r}^\pm
\,,&\qquad&[H_m,G_r^\pm]&=&\pm G_{m+r}^\pm\,,
\\
\L[L_m,H_n\R]&=&{}-nH_{m+n}\\
\L\{G_r^-,G_s^+\R\}&=&\multicolumn{5}{l}{2L_{r+s}-(r-s)H_{r+s}+
{\ctop\over3}(r^2-\frac{1}{4})
\Kr{r+s}\,,}\EA\label{N2algebra}
\EE

\noi
where the fermionic modes take half-integer values.
The question naturally arises now whether or not the topological singular 
vectors are also a rewriting of the NS singular vectors. For the case
of singular vectors built on chiral primaries the answer is 
that {\it only the} \Gn{\it -closed topological singular vectors
transform into NS singular vectors after the untwisting}. Moreover,
the resulting NS singular vectors
are built on chiral primaries of the NS algebra
when using the twist $T_{W2}$ \req{twb},
or on antichiral primaries when using the twist $T_{W1}$ \req{twa}.
On the other hand, {\it the twisting of NS singular vectors always
produces topological singular vectors annihilated by} \Gn. They are built
on chiral topological primaries provided the NS singular vectors
are built on chiral or antichiral primaries; otherwise, the 
topological singular vectors would be built on topological \Gn-closed
primaries which are not chiral.

All these statements can be verified rather easily. One only
needs to investigate how the highest weight (h.w.) conditions
satisfied by the singular vectors get modified under the
twistings or untwistings given by \req{twa} and \req{twb}.
By inspecting these it is obvious that the bosonic h.w.
conditions, \ie\ $L_{m>0}\ket{\chi_{NS}}=H_{m>0}\ket{\chi_{NS}}=0$,
on the one hand, and $\cL_{m>0}\ket{\chi_T}=\cH_{m>0}\ket{\chi_T}=0$,
on the other hand, are conserved under the twistings and untwistings.
In other words, if the topological secondary state
 $\ket{\chi_T}$ satisfies the topological
bosonic h.w. conditions, then the corresponding untwisted secondary state
$\ket{\chi_{NS}}$ satisfies the NS bosonic h.w. conditions
and vice-versa. With the fermionic h.w. conditions things are
not so straightforward. While the h.w. conditions 
$\cQ_{m>0}\ket{\chi_T}=0$ are converted into h.w. conditions of the
type $G_{m\geq\half}^{\pm}\ket{\chi_{NS}}=0$
($G^+$ or $G^-$ depending on the specific twist), in both twists \Gn\
is transformed into one of the $G_{1/2}^{\pm}$ modes, as we pointed out
in last section. But
$G_{1/2}^{\pm}\ket{\chi_{NS}}=0$ is nothing but a h.w. condition satisfied 
by all the NS singular vectors!
As a result, the twisting of a NS singular vector always
produces a topological singular vector annihilated by \Gn. In addition,
the BRST-invariance condition $\Qz\ket{\P}=0$ on the chiral topological
primaries is transformed into the antichirality ($G_{-1/2}^-\ket{\P}=0$) 
and the chirality ($G_{-1/2}^+\ket{\P}=0$) conditions on the NS 
primary states, under $T_{W1}$ and $T_{W2}$ respectively.

Now let us analyze the transformation of the $(\cL_0, \cH_0)$
eigenvalues $(l, q+ \htop )$ of the \Gn-closed topological singular vectors
$\ket{\chi_T}_l^{(q)G}$ into $(L_0, H_0)$ eigenvalues
$(\Delta' + l', q'+ \htop' )$ of the NS singular vectors
$\ket{\chi_{NS}}_l^{(q)}$. In other words, given a \Gn-closed topological
singular vector in the chiral Verma module $V_T(\htop)$, at level 
$l$ and with relative charge $q$, let us determine the NS Verma 
module $V_{NS}(\D',\htop')$, the level $l'$ and the relative charge 
$q'$ of the corresponding
untwisted NS singular vector. Observe that $\D'$ denotes the conformal
weight of the primary on which the NS singular vector is built
($\D=0$ for the chiral topological primaries as we discussed before).
Using $T_{W1}$ \req{twa} the U(1) charge does not change, $\htop'=\htop$,
whereas $\D'=-\htop/2=-\htop'/2$ and the level gets modified as
$l'=l- \half q$. Therefore the \Gn-closed topological 
singular vectors of the types $\ket{\chi_T}_l^{(0)G}$
and $\ket{\chi_T}_l^{(1)G}$, in chiral topological Verma 
modules, are transformed under $T_{W1}$ into NS singular vectors
of the types $\ket{\chi_{NS}}_l^{(0)a}$ and
$\ket{\chi_{NS}}_{l-\half}^{(1)a}$ in antichiral NS Verma modules,
as indicated by the superscript $a$.
Using $T_{W2}$ \req{twb}, on the other hand, the U(1)
charge reverses its sign, $\htop'=-\htop$
while $\D'=-\htop/2=\htop'/2$ and the level
gets modified, again as $l'=l - \half q$. Therefore the 
\Gn-closed topological singular vectors of
the types $\ket{\chi_T}_l^{(0)G}$
and $\ket{\chi_T}_l^{(1)G}$, in chiral topological Verma modules, 
 result under $T_{W2}$ 
in NS singular vectors of the types $\ket{\chi_{NS}}_l^{(0)ch}$ and
$\ket{\chi_{NS}}_{l-\half}^{(-1)ch}$ in chiral NS Verma modules,
as indicated by the superscript $ch$. Notice that
the values of $\D'$, in terms of $\htop$, are the same under the two 
untwistings.

An interesting result here is that the same charged \Gn-closed topological
singular vector $\ket{\chi_T}_l^{(1)G}$, at level $l$, 
gives rise to both charge $q=1$
and charge $q=-1$ NS singular vectors at level $l-\half$, built
on antichiral primaries and chiral primaries respectively.
Furthermore, since there are no other types of charged \Gn-closed
topological singular vectors in chiral Verma modules, we deduce that
{\it all the charged
NS singular vectors built on chiral primaries have relative 
charge} $q=-1$, {\it whereas all the charged NS singular
vectors built on antichiral primaries have relative
charge} $q=1$. Moreover, the $q=1$ and $q=-1$
singular vectors are mirrored under the interchange
$H_m \leftrightarrow - H_m$ (therefore $\htop \leftrightarrow
-\htop$) and $G_r^+ \leftrightarrow G_r^-$.

Finally, an important result is the fact that the charged and uncharged
NS singular vectors $\ket{\chi_{NS}}_{l}^{(0)a}$
and $\ket{\chi_{NS}}_{l-\half}^{(1)a}$, on the one hand, and
$\ket{\chi_{NS}}_{l}^{(0)ch}$
and $\ket{\chi_{NS}}_{l-\half}^{(-1)ch}$, on the other hand,
 must come in pairs, although
in different Verma modules, since they are just the untwistings
of the \Gn-closed topological singular vectors $\ket{\chi_T}_l^{(0)G}$ and 
$\ket{\chi_T}_l^{(1)G}$,
which come in pairs inside the four-member topological families
described in the previous subsection. Namely, from the diagram \req{dia}
one deduces the one-to-one mapping
$\ket{\chi_T}_{l,-\htop-\ctop/3}^{(1)G}=
\cG_0\,\cA\ket{\chi_T}_{l,\,\htop}^{(0)G}$
between uncharged and charged $\cG_0$-closed topological singular 
vectors in the chiral topological Verma modules $V_T(\htop)$ and
$V_T(-\htop-\ctop/3)$ respectively. Untwisting this mapping using $T_{W1}$ 
\req{twa} and $T_{W2}$ \req{twb} one obtains the one-to-one mappings

\BE
\ket{\chi_{NS}}_{l-{1\over2},-\htop-{\ctop\over3}}^{(1)a}=
G_{1/2}^+\cA_1\,\ket{\chi_{NS}}_{l,\,\htop}^{(0)a}{\ },{\ \ \ \ }
\ket{\chi_{NS}}_{l-{1\over2},\,\htop+{\ctop\over3}}^{(-1)ch}=
G_{1/2}^-\cA_{-1}\,\ket{\chi_{NS}}_{l,-\htop}^{(0)ch}
\label{svcp}
\EE

\noi
between uncharged and charged NS singular vectors in the antichiral NS
Verma modules $V_{NS}^a(\htop)$ and $V_{NS}^a(-\htop-\ctop/3)$, on the
left-hand side, and in the chiral NS Verma modules $V_{NS}^{ch}(-\htop)$ and
$V_{NS}^{ch}(\htop+\ctop/3)$, on the right-hand side. $\cA_1$ and
$\cA_{-1}$ denote the untwisted spectral flow $\cA_{\th}$ 
\cite{BJI4}, which will be given in eq. \req{ospfl}, for the values 
$\th=\pm 1$.

\section{Spectrum of Topological and NS Singular Vectors in Chiral
Verma Modules. Subsingular Vectors}\lvm

We have just shown that the spectrum of U(1) charges corresponding
to the chiral Verma modules which contain
topological and NS singular vectors
is the same for the singular vectors of the types $\ket{\chi_T}_l^{(0)G}$, 
 and $\ket{\chi_{NS}}_l^{(0)a}$, on the one hand,
and also the same for the singular vectors of the
types $\ket{\chi_T}_l^{(1)G}$ and
 $\ket{\chi_{NS}}_{l-\half}^{(1)a}$, on the other hand,
 where the superscript $a$
indicates that the NS vectors are built on antichiral
primaries. Let us introduce $\htop^{(0)}$ and $\htop^{(1)}$ such 
that the first spectrum consists of all possible values of
 $\htop^{(0)}$ whereas the second spectrum consists of all 
 possible values of $\htop^{(1)}$, these values being connected
to each other by the relation  $\htop^{(1)} =
-\htop^{(0)} - {\ctop\over3}$. The spectrum of U(1) charges
for the case of NS Verma modules built on chiral
 primaries, in turn, is the corresponding to $(-\htop^{(0)})$,
for singular vectors of the type $\ket{\chi_{NS}}_l^{(0)ch}$,
and to $(-\htop^{(1)})$, for singular
vectors of the type $\ket{\chi_{NS}}_{l-\half}^{(-1)ch}$.

In what follows we will derive a conjecture for the 
spectrum of possible values of $\htop^{(0)}$ and $\htop^{(1)}$.
These values can be viewed as the roots of the chiral determinants
for the Topological algebra, as well as the roots of
the antichiral determinants for the NS algebra,
whereas $(-\htop^{(0)})$ and $(-\htop^{(1)})$ are the roots of the
chiral determinants for the NS algebra. In the case that the values of
$\htop^{(0)}$ and/or $\htop^{(1)}$ predict singular vectors which do
not exist in the complete Verma modules, \ie\, only in the chiral
Verma modules, then one has encountered subsingular vectors.

\subsection{Spectrum of Singular Vectors}\lvm

Let us concentrate on the  NS algebra for convenience.
In principle we cannot compute the spectra $\htop^{(1)}$ and
 $\htop^{(0)}$, corresponding to charged and uncharged NS singular vectors 
built on antichiral primaries, simply by imposing 
the relation $\Delta = - {\htop \over2}$ in the spectra given
 by the roots of the determinant formula for the  NS algebra 
\cite{BFK},\cite{Nam},\cite{KaMa3}. 
The reason is that the NS determinant formula does not apply to
 incomplete Verma modules constructed on chiral or antichiral
 primary states annihilated by $G_{-\half}^+$ or $G_{-\half}^-$
 (for which, as a result, $\Delta ={\htop \over2}$ or 
$\Delta = - {\htop \over2}$ respectively, but not the other way around).
Nevertheless let us start our analysis with the following ansatz. 

{\it Ansatz}.$-$ 
The sets of roots of the chiral and antichiral NS determinants  
coincide with the sets of roots of the NS determinants
for the particular cases $\Delta =\pm{\htop \over2}$, respectively.

Our strategy will be now to analyze the set of roots of the NS 
determinant formula
for $\D=\pm{\htop\over2}$ taking into account that, as we deduced in
the last section, the charged and uncharged NS singular vectors 
built on chiral primaries come in pairs, with a precise relation between 
their corresponding Verma modules. Namely, for singular vectors in 
antichiral Verma modules, satisfying $\D=-\htop/2$, the relation is
$\htop^{(1)}=-\htop^{(0)}-{\ctop\over3}$, while for 
singular vectors in chiral Verma modules, satisfying $\D=\htop/2$, it is
 $\htop^{(1)}=-\htop^{(0)}+{\ctop\over3}$
 (in this last expression $\htop^{(1)}$ 
and $\htop^{(0)}$ denote the true spectra 
 for the chiral Verma modules, instead of ($-\htop^{(1)}$) and 
($-\htop^{(0)}$) in our previous notation).

The roots of 
the determinant formula for the  NS algebra
\cite{BFK},\cite{Nam},\cite{KaMa3}
are given by the solutions of the quadratic 
vanishing surface $f_{rs}^A=0$, with

\BE   f_{r,s}^A = 2 \L({\ctop-3\over3}\R) \Delta - \htop^2
 -{1\over4} \L({\ctop-3\over3}\R)^2 +
 {1\over4} \L(\L({\ctop-3\over3}\R) r + s \R)^2 \qquad r\in\oZ^+\,,\,\,
  s\in2\oZ^+ \label{frs} \EE

\noi
and the solutions of the vanishing plane $g_k^A = 0$,
with
       
\BE   g_k^A = 
  2 \Delta-2k\htop + \L({\ctop-3\over3}\R)(k^2-{1\over4})
\qquad k\in\oZ+\half \label{gk} \EE

\noi
In the complete Verma modules the solutions to $f_{rs}^A=0$ and 
$g_k^A = 0$ correspond to uncharged and charged singular vectors,
respectively. 

Let us concentrate on the antichiral case, for convenience.
Solving for $f_{r,s}^A = 0$, with $\Delta = - {\htop\over2}$, one finds
two two-parameter solutions for $\htop$ (since
$f_{r,s}^A = 0$ becomes a quadratic equation for $\htop$).
These solutions are

\BE  \htop_{r,s} = - \half \L(
      \L({\ctop-3\over3}\R) (r+1) + s \R)  \label{hrs} \EE
\noi 
and

\BE  \hat\htop_{r,s} = \half \L(
      \L({\ctop-3\over3}\R) (r-1) + s \R)\,.  \label{hh0rs} \EE

\noi                                                   
Solving for $g_k^A = 0$, with $\Delta = - {\htop\over2}$,
 one finds the one-parameter solution

\BE \htop_k =  \L({\ctop-3\over6}\R) (k-\half)\,,
                                          \label{hk}\EE

\noi
except for $k=-\half$ where $g_k^A$ is identically zero (\ie\ all the
states $G^-_{1/2}\ket{\D,\htop}$ with $\D=-{\htop\over2}$ are 
singular vectors).

If our ansatz is correct, the set of roots of the antichiral NS
determinant formula should be equal as the set of roots
given by the solutions \req{hrs}, \req{hh0rs} and \req{hk}.
The problem at hand is
therefore to distribute all these solutions into two sets,
say $H^{(0)}$ and $H^{(1)}$, such that for any given
solution $\htop^{(0)}$ in the set $H^{(0)}$
there exists one solution $\htop^{(1)}$ in the set $H^{(1)}$,
satisfying $\htop^{(0)} = -\htop^{(1)} - {\ctop\over3}$,
and vice-versa. For this purpose it is helpful to write
down the expressions corresponding to
$(-\htop_{r,s} - {\ctop\over3})$,
 ${\ }(-\hat\htop_{r,s} - {\ctop\over3}){\ }$
and ${\ }(-\htop_k - {\ctop\over3})$. These are given by

\BE -\htop_{r,s} - {\ctop\over3} {\ } = {\ } \half \L(
      \L({\ctop-3\over3}\R) (r-1) + s - 2 \R)\,,  \label{mh0rs} \EE

\BE -\hat\htop_{r,s} - {\ctop\over3} {\ } = - \half \L(
      \L({\ctop-3\over3}\R) (r+1) + s + 2 \R)  \label{mhh0rs} \EE

\noi
and

\BE -\htop_k - {\ctop\over3} {\ } = - \half \L(
      \L({\ctop-3\over3}\R) (k+{3\over2}) + 2 \R)\,.  \label{mhk} \EE

Comparing these expressions with the set of solutions
$\htop_{r,s}{\ }$, $\hat\htop_{r,s}{\ }$ and ${\ }\htop_k{\ }$, given by
\req{hrs}, \req{hh0rs} and \req{hk}, one finds straightforwardly

\BE   \htop_{r,2} = - \htop_{r-\half} - {\ctop\over3}\,, \qquad
    \htop_{r,s>2} = - \hat\htop_{r,(s-2)} - {\ctop\over3}\,.
\EE

\noi
Therefore the simplest solution to the problem is that the
spectrum corresponding to the uncharged singular vectors
$\ket{\chi_{NS}}_l^{(0)a}$ is given by
${\ }\htop_{r,s}^{(0)} = \htop_{r,s}{\ }$, 
eq. \req{hrs}, with the level of the
state $l = {rs\over2}$ as usual \cite{BFK}, whereas the
spectrum corresponding to the charged
singular vectors $\ket{\chi_{NS}}_{l-1/2}^{(1)a}$, at level
$l-\half={rs-1\over2}$, 
 is given by 
${\ }\htop_{r,s}^{(1)} = - \htop_{r,s} - {\ctop\over3}{\ }$, that is

\BE  \htop_{r,s}^{(0)} = - \half \L(
\L({\ctop-3\over3}\R) (r+1) + s \R) \ \ r\in \oZ^+, \ \ s\in 2\oZ^+, 
\label{h0rs} \EE

\noi
and

\BE  \htop_{r,s}^{(1)} = \half \L(
 \L({\ctop-3\over3}\R) (r-1) + s - 2 \R)  \ \ r\in \oZ^+, \ \ s\in 2\oZ^+,
      \label{h1rs} \EE
 
\noi
with $\htop_{r,s}^{(1)}$ containing the two series  $\htop_k$ and
$\hat\htop_{r,s}{\ }$. Namely, for $s=2{\ }$
$\htop_{r,2}^{(1)} = \htop_{r-\half}$, where $l-\half=r-\half$ is
the level of the charged singular vector, while for $s>2{\ }$
$\htop_{r,s>2}^{(1)} = \hat\htop_{r,(s-2)}$, with the level
given by $l-\half = {rs-1 \over2}{\ }$. This implies the existence of
{\it charged subsingular} vectors in the complete Verma modules with
$\D=-\htop/2$, $\htop=\htop_{r,s>2}^{(1)}$, since these 
Verma modules contain
uncharged singular vectors rather than charged. The complete Verma modules
with  $\D=-\htop/2$, $\htop=\htop_{r,2}^{(1)}$, however, contain
charged singular vectors like the antichiral Verma modules.

We see therefore that for this solution half of the zeroes of the 
quadratic vanishing surface $f_{r,s}^A = 0$, with $\D=-\htop/2$, for 
every pair $(r,s)$,
correspond to uncharged singular vectors $\ket{\chi_{NS}}_l^{(0)a}$ at
level $l={rs\over2}$,
and the other half correspond to charged singular vectors
$\ket{\chi_{NS}}_{l-1/2}^{(1)a}$, which are subsingular in the complete 
Verma modules, at
level $l-\half = {r(s+2)-1 \over2}{\ }$. The zeroes of the vanishing 
plane $g_k^A = 0$, for $\D=-\htop/2$, in turn, correspond to charged 
singular vectors $\ket{\chi_{NS}}_{l-1/2}^{(1)a}$ at level $l-\half=k$,
with $k>0$. 

Before searching for more intricate solutions
let us have a look at the data for $\htop^{(0)}$ and
$\htop^{(1)}$ given by the singular vectors
themselves. We have computed until level 3 all the topological 
singular vectors in chiral topological Verma modules and all
the NS singular vectors in chiral and antichiral NS Verma modules.
The explicit expressions for the topological singular vectors we
have given recently in \cite{BJI6} 
(some of those singular vectors were already published). 
The explicit expressions for the NS
singular vectors will be given in \cite{BJI8}, although in the Appendix
we also write down
and analyze the NS singular vectors $\ket{\chi_{NS}}_{1}^{(0)}$, 
$\ket{\chi_{NS}}_{3\over2}^{(1)}$ and $\ket{\chi_{NS}}_{3\over2}^{(-1)}$
in chiral, antichiral and complete Verma modules.
 Here we need only the values of $\htop^{(0)}$ and $\htop^{(1)}$ .
These are the following:
\vskip .2in

- For $\ket{\chi_{NS}}_1^{(0)a}$, 
$\ket{\chi_T}_1^{(0)G}{\ } {\rm and} {\ } \ket{\chi_T}_1^{(-1)Q}$
$ {\ \ } \htop^{(0)}= -{\ctop\over3}$

\vskip .17in

- For $\ket{\chi_{NS}}_{\half}^{(1)a}$,
$\ket{\chi_T}_1^{(1)G}{\ } {\rm and} {\ } \ket{\chi_T}_1^{(0)Q}$
${\ \ } \htop^{(1)}= 0$ 

\vskip .17in

- For $\ket{\chi_{NS}}_2^{(0)a}$, 
$\ket{\chi_T}_2^{(0)G}{\ } {\rm and} {\ } \ket{\chi_T}_2^{(-1)Q}$ 
$ {\ \ } \htop^{(0)}=
 {1-\ctop\over2}{\ },{\ \ } -{\ctop+3\over3} $

\vskip .17in

- For $\ket{\chi_{NS}}_{3\over2}^{(1)a}$,
$\ket{\chi_T}_2^{(1)G}{\ } {\rm and} {\ }\ket{\chi_T}_2^{(0)Q} $
${\ \ } \htop^{(1)}={\ctop-3\over 6}{\ }, 
{\ \ }1$                                                 

\vskip .17in

- For $\ket{\chi_{NS}}_3^{(0)a}$, 
$\ket{\chi_T}_3^{(0)G}{\ } {\rm and} {\ }\ket{\chi_T}_3^{(-1)Q}$
${\ \ }\htop^{(0)}={3-2\ctop\over3}{\ },{\ \ }
-{\ctop+6\over3}$

\vskip .17in

- For $\ket{\chi_{NS}}_{5\over2}^{(1)a}$, 
$\ket{\chi_T}_3^{(1)G}{\ } {\rm and} {\ }\ket{\chi_T}_3^{(0)Q}$
${\ \ }\htop^{(1)}={\ctop-3\over3}{\ },
{\ \ }2 $

\vskip .2in

\noi
(we remind that the topological singular vectors of types
$\ket{\chi_T}^{(1)G}$ and $\ket{\chi_T}^{(0)Q}$
are together in the same Verma module, at the same level). 
In addition, the BRST-invariant uncharged topological
singular vector at level 4, \ie\ $\ket{\chi_T}_4^{(0)Q}$, has been
computed in \cite{Sem} with the result
${\ }\htop^{(1)} = {\ctop-3\over2}, {\ }{\ctop+3\over6}, {\ }3$.
These values also correspond to the singular vectors 
${\ }\ket{\chi_T}_4^{(1)G}{\ }$ and 
$\ket{\chi_{NS}}_{7\over2}^{(1)a}$, as we have deduced.

By comparing these results with the roots of the NS
determinant formula for $\D=-\htop/2$, given by
expressions \req{hrs}, \req{hh0rs} and \req{hk},
 we notice that the values
we have found for $\htop^{(0)}$ fit exactly in the 
expression $\htop_{r,s}$ \req{hrs}, that is, the upper
solution for the quadratic vanishing surface,
but not in the lower solution $\hat\htop_{r,s}$. The values we
have found for $\htop^{(1)}$ follow exactly the prediction of
the spectral flow automorphism for each case, \ie\
${\ }\htop^{(1)} = -\htop^{(0)} - {\ctop\over3}{\ }$.
Hence the actual spectrum of U(1) charges $\htop^{(0)}$ and 
$\htop^{(1)}$, as far as we can tell, follows 
the pattern we have found under the ansatz that the set of roots of the 
antichiral NS determinant formula coincide with the set of roots
of the NS determinant formula for $\Delta = -{\htop\over2}$.
That is, $\htop_{r,s}^{(0)} = \htop_{r,s}$  is
given by \req{h0rs} and  
 $\htop_{r,s}^{(1)} = - \htop_{r,s} - {\ctop\over3}$ is given
by \req{h1rs}\footnote{The spectrum $\htop_{r,s}^{(1)}$, eq.
\req{h1rs}, was written for 
the first time in ref. \cite{Sem}, eq. (3.1), (with $s \in \oZ^+$)
fitting the data obtained from  
the computation of the topological singular vectors of type
$\ket{\chi_T}^{(0)Q}$, referred simply as ``the topological singular
vectors".
However this spectrum was never related to
charged singular vectors of the NS algebra, the author believed 
that it was related to uncharged NS singular vectors instead,  
and was given directly by the uncharged roots of the
NS determinant formula for the case $\Delta = -{\htop\over2}$. 
The fact that this spectrum corresponds to
{\it charged singular} vectors in chiral NS Verma modules is precisely the
reason why it corresponds, for the values $\htop_{r,s>2}^{(1)}$,
to {\it charged subsingular} vectors in the complete Verma modules.}.

For the case of NS singular vectors in chiral Verma modules 
one finds the same values for the U(1) charges as for the case of
antichiral Verma modules but with the sign reversed, as expected.
 Therefore the spectra of U(1) charges for the
NS singular vectors of the types $\ket{\chi_{NS}}_{l}^{(0)ch}$
and $\ket{\chi_{NS}}_{l-\half}^{(-1)ch}$ are given by 
$(-\htop^{(0)}_{r,s})$ and $(-\htop^{(1)}_{r,s})$ respectively.

\vskip .2in

{\it Summary of Results}.$-$
Let us summarize our results for the spectra of U(1) charges
and conformal weights corresponding to the Verma modules
 which contain the different kinds of singular vectors. Although
 these results have been checked only until level 4 we conjecture
 that they hold at any level. We remind that $\htop_{r,s}^{(0)}$ 
 is given by \req{h0rs} and $\htop_{r,s}^{(1)} = 
 - \htop_{r,s}^{(0)} - {\ctop\over3}$ is given by \req{h1rs}.

\vskip .17in

1) The spectrum of U(1) charges corresponding to the chiral 
topological Verma modules which contain
singular vectors of the types $\ket{\chi_T}_l^{(0)G}$ and $
\ket{\chi_T}_l^{(-1)Q}$,
on the one hand, and singular vectors of the types $\ket{\chi_T}_l^{(1)G}$
and $\ket{\chi_T}_l^{(0)Q}$ , on the other hand,
is given by ${\ }\htop^{(0)}_{r,s}{\ }$, and ${\ }\htop^{(1)}_{r,s}{\ }$,
respectively, with $l = {r s \over 2}$. These are the only types of
topological singular vectors which exist in chiral topological
Verma modules.
 
\vskip .17in

2) The spectra of U(1) charges and conformal weights corresponding to 
antichiral NS Verma modules which contain uncharged
and charged NS singular vectors, $\ket{\chi_{NS}}_{l}^{(0)a}$ 
and $\ket{\chi_{NS}}_{l-\half}^{(1)a}$, are given by $\htop^{(0)}_{r,s}$,
$\Delta^{(0)}_{r,s} = -\half \htop^{(0)}_{r,s}$, and
 $\htop^{(1)}_{r,s}$, 
 $\Delta^{(1)}_{r,s} = -\half \htop^{(1)}_{r,s}$, respectively,
  with $l = {r s \over 2}$. These are the only types of NS singular 
  vectors which exist in antichiral NS Verma modules.
 
\vskip .17in

3) The spectra of U(1) charges and conformal weights 
corresponding to chiral NS Verma modules which contain uncharged and
charged NS singular vectors, $\ket{\chi_{NS}}_{l}^{(0)ch}$ and
$\ket{\chi_{NS}}_{l-\half}^{(-1)ch}$, are given by $(-\htop^{(0)}_{r,s})$,  
$\Delta^{(0)}_{r,s} = \half (-\htop^{(0)}_{r,s})$, 
and $(-\htop^{(1)}_{r,s})$, 
$\Delta^{(1)}_{r,s} = \half (-\htop^{(1)}_{r,s})$, respectively, 
with $l = {r s \over 2}$. These are the only types of NS singular
vectors which exist in chiral NS Verma modules.
Therefore the spectrum of conformal weights is the same for 
chiral NS Verma modules as for antichiral NS Verma modules,
although the spectrum of U(1) charges, as well as the relative charge
of the singular vectors, is reversed in sign.

\subsection{Subsingular Vectors}\lvm

Subsingular vectors are singular vectors only in the quotient of 
complete Verma modules by submodules generated by singular vectors.
This is indeed the case at hand because chiral and antichiral NS Verma
modules are nothing but the quotient of complete NS Verma modules
by the singular vectors $G_{-1/2}^+\ket{\D,\htop}$, $\D=\htop/2$, and
$G_{-1/2}^-\ket{\D,\htop}$, $\D=-\htop/2$, respectively.
Similarly, the chiral topological Verma modules are the quotient of
complete topological Verma modules of types $V_T(\ket{0,\htop}^Q)$
and $V_T(\ket{0,\htop}^G)$, by the submodules generated by the
singular vectors $\Gz\ket{0,\htop}$ and 
$\Qz\ket{0,\htop}$, respectively (in ref. \cite{BJI6} there is a
detailed description of the complete topological Verma modules and
the spectra of conformal weights and U(1) charges corresponding
to singular vectors). 

We have found the remarkable fact that (at least until level 4) 
half of the uncharged NS singular
vectors with levels $l={rs\over2}$ in the complete Verma modules
with $\D=\mp {\htop\over2}$, $\htop = \pm \hat\htop_{r,s}$, 
eq. \req{hh0rs}, or equivalently $\htop = \pm \htop_{r,(s+2)}^{(1)}$,
eq. \req{h1rs},
 have been replaced by charged NS singular vectors with levels
$l-\half = {r(s+2)-1\over2}$, in the antichiral and chiral Verma 
modules, respectively. These charged singular vectors are therefore
charged subsingular vectors in the complete Verma modules. Since
$s\in2\oZ^+$, the complete Verma modules with $\D=\mp{\htop\over2}$
which contain charged subsingular vectors are those with
$\htop=\pm\htop_{r,s>2}^{(1)}$ (in other words, the set of U(1)
charges given by $\htop_{r,s>2}^{(1)}$ is equal to the set given by 
$\hat\htop_{r,s}$, as we deduced in the previous subsection), whereas
the complete Verma modules with $\htop=\pm\htop_{r,2}^{(1)}$ do not
contain charged subsingular vectors, but charged singular vectors
instead (which are singular both in the complete Verma modules and
in the chiral Verma modules).

In the case of the Topological algebra there is a much more
symmetrical situation with respect to charged and uncharged
singular and subsingular vectors. Namely, there are charged
as well as uncharged subsingular vectors, and charged as well
as uncharged singular vectors which vanish in the chiral Verma modules.
One finds\footnote{The
subsingular vectors of the Topological algebra were not considered
in the first version of this paper. Here we borrow our own results 
from ref. \cite{BJI6}, where we have published all the topological
subsingular vectors at levels 2 and 3 which become singular in
chiral topological Verma modules.} that
$\htop=\htop_{r,s>2}^{(1)}$ corresponds to charged and uncharged
subsingular vectors of types
$\ket{\chi_T}_l^{(1)G}$ and $\ket{\chi_T}_l^{(0)Q}$, 
in the complete Verma modules $V_T(\ket{0,\htop}^G)$, whereas
$\htop=\htop_{r,s>2}^{(0)}$ corresponds to charged and uncharged
subsingular vectors of types
$\ket{\chi_T}_l^{(0)G}$ and $\ket{\chi_T}_l^{(-1)Q}$
in the complete Verma modules $V_T(\ket{0,\htop}^Q)$. All of these 
topological subsingular vectors become singular in the chiral
topological Verma modules $V_T(\ket{0,\htop}^{G,Q})$.

Under the untwistings $T_{W1}$ and $T_{W2}$ only the $\cG_0$-closed 
topological
h.w. states (primary and singular vectors) remain h.w. states of the
NS algebra. Therefore the complete topological Verma modules of type
$V(\ket{0,\htop}^Q)$ are not transformed into NS Verma modules, 
and the BRST-invariant uncharged topological singular
vectors $\ket{\chi_T}_l^{(0)Q}$, in the chiral topological Verma modules,
are not transformed into singular vectors in the chiral or antichiral
NS Verma modules. In this manner the symmetry between
charged and uncharged topological subsingular vectors is broken under the
untwistings, and one only finds charged subsingular vectors in
the NS algebra, for Verma modules with $\D=\mp{\htop\over2}$,
$\htop=\mp\htop_{r,s>2}^{(1)}$.

In the Appendix we
analyze the complete situation for the particular case of the uncharged
singular vectors  $\ket{\chi_{NS}}_1^{(0)}$ and the
charged singular vectors $\ket{\chi_{NS}}_{3\over2}^{(1)}$ and
 $\ket{\chi_{NS}}_{3\over2}^{(-1)}$.
We write down the h.w. equations, with their solutions, for
the primary states being non-chiral, chiral and antichiral.
We show that the uncharged singular
vectors $\ket{\chi_{NS}}_1^{(0)}$, for $\Delta = \mp {\htop\over2}{\ },$
 $\htop = \pm \hat\htop_{1,2}= \pm 1{\ }$, or equivalently 
 $\htop = \pm \htop_{1,4}^{(1)}= \pm 1{\ },$
 vanish once we switch on
antichirality and chirality on the primary states, respectively,
whereas the charged subsingular vectors become the
charged singular vectors 
$\ket{\chi_{NS}}_{3\over2}^{(1)a}$ with $\htop=1$, and 
$\ket{\chi_{NS}}_{3\over2}^{(-1)ch}$ with $\htop=-1$. 
We also show that the
subsingular vectors are non-highest
weight null vectors, after we switch off antichirality and chirality
respectively on the primary states, which are not
descendant states of any singular vectors, as expected (they can descend
down to a singular vector but not the other way around). 

\vskip .16in

A final remark is that subsingular vectors do not
exist for Verma modules of the Virasoro 
algebra neither for Verma modules of the sl(2) algebra. 
The very existence of these objects for the N=2 Superconformal algebras 
has been so far unknown\footnote{unknown until the first version 
of this paper appeared in hep-th/9602166 (1996)},
in spite of the fact that the issue has been 
investigated during the last few years \cite{KD}. Observe that the 
subsingular vectors that we have found are those which become
singular in chiral Verma modules. The issue whether or not these are
the only subsingular vectors of the Topological and NS algebras 
is currently under investigation.

\section{Spectrum of R Singular Vectors on Ramond Ground States.
Subsingular Vectors}\lvm

The singular vectors of the NS algebra transform into
singular vectors of the R algebra under the action of
the spectral flows, and vice-versa 
\cite{SS},\cite{LVW},\cite{Kir1},\cite{BJI4}. 
In particular, the NS singular vectors built
on chiral or antichiral primaries transform into R singular vectors
built on the Ramond ground states. As a consequence, we can
write down the spectrum of U(1) charges corresponding to the R
Verma modules built on
the R ground states which contain singular vectors, simply by applying
the spectral flow
transformations to the spectra \req{h0rs} and \req{h1rs}
found in last section. Before doing this let us say a few
words about the R algebra.

\vskip .17in

{\it The R Algebra}.$-$ The Periodic N=2 Superconformal algebra 
is given by \req{N2algebra}, where the fermionic generators
$G^{\pm}_r$ are integer moded. The zero modes of the
fermionic generators characterize the states as being
$G_0^+$-closed or $G_0^-$-closed, as the anticommutator
$\{G_0^+,G_0^-\} = 2L_0 - {\ctop\over12}{\ }$ shows.
The R ground states are annihilated by both $G_0^+$
and $G_0^-$, therefore the conformal weight satisfies
$\Delta = {\ctop\over24}$ for them, as a result.

In order to simplify the analysis that follows it is very
convenient to define the U(1) charge for the states of the
R algebra in the same way as for the states of the
NS algebra. Namely, the U(1) charge of the states
will be denoted by $\htop$, instead of $\htop \pm \half$,
whereas the relative charge $q$ of a secondary state will
be defined as the difference between the $H_0$-eigenvalue
of the state and the $H_0$-eigenvalue of the primary on which it
is built. Therefore, the relative charges of the R states
are defined to be integer, in contrast with the usual definition.
We will denote the R singular vectors as $\ket{\chi_R}_l^{(q)+}$ and
$\ket{\chi_R}_l^{(q)-}$, where, in addition to the level and the 
relative charge, we indicate that the vector is annihilated by
$G_0^+$ or $G_0^-$.  

\vskip .17in
{\it The Spectral Flows}. $-$
The ``usual" spectral flow \cite{SS}, \cite{LVW}, is given by the
one-parameter family of transformations

\BE\new\BA{rclcrcl}
\cU_\th \, L_m \, \cU_\th^{-1}&=& L_m
 +\th H_m + {\ctop\over 6} \th^2 \delta_{m,0}\,,\\
\cU_\th \, H_m \, \cU_\th^{-1}&=&H_m + {\ctop\over3} \th \delta_{m,0}\,,\\
\cU_\th \,\, G^+_r \, \cU_\th^{-1}&=&G_{r+\th}^+\,,\\
\cU_\th \,\, G^-_r \, \cU_\th^{-1}&=&G_{r-\th}^-\,,\
\label{spfl} \EA\EE

\noi
satisfying $\cU_{\th}^{-1} = \cU_{(-\th)}$ and giving rise to
isomorphic algebras. If we denote by $(\Delta, \htop)$ the 
$(L_0, H_0)$ eigenvalues of any given state, then 
the eigenvalues of the transformed state
$\cU_{\th} \kc$ are
 $(\Delta-\th \htop +{c\over6} \th^2, \htop - {c\over3} \th)$.
If the state $\kc$ is a level-$l$ secondary state with relative
charge $q$, then one gets straightforwardly
that the level of the transformed state $\cU_{\th} \kc$ changes
to $l-\th q$, while the relative charge remains equal. 

There is another spectral flow \cite{BJI4}, \cite{BJI7}, mirrored to
the previous one, which is the untwisting of the 
Topological algebra automorphism \req{autom} (for general values 
of $\th$), given by

\BE\new\BA{rclcrcl}
\cA_\th \, L_m \, \cA_\th&=& L_m
 +\th H_m + {\ctop\over 6} \th^2 \delta_{m,0}\,,\\
\cA_\th \, H_m \, \cA_\th&=&- H_m - {\ctop\over3} \th \delta_{m,0}\,,\\
\cA_\th \, G^+_r \, \cA_\th&=&G_{r-\th}^-\,,\\
\cA_\th \, G^-_r \, \cA_\th&=&G_{r+\th}^+\,.\
\label{ospfl} \EA\EE

\noi
with $\cA_{\th}^{-1} = \cA_{\th}$.
The $(L_0, H_0)$ eigenvalues of the transformed states 
$\cA_{\th} \kc$ are now
 $(\Delta+\th \htop +{c\over6} \th^2, - \htop - {c\over3} \th)$
(that is, they differ from the previous case by the
 interchange $\htop \rightarrow -\htop$). From this one easily
deduces that, under the spectral flow $\cA_{\th}$, the level $l$
of any secondary state changes to $l + \th q$ while  
the relative charge $q$ reverses its sign.

For half-integer values of $\theta$ the two spectral flows
interpolate between the NS algebra and the 
R algebra. In particular, for $\theta = \half$ the h.w. states
of the NS algebra (primaries as well as singular vectors) become h.w. states
of the R algebra with helicity $(-)$ (\ie\ annihilated by
$G_0^-$), while for $\theta = -\half$ the h.w. states of the NS
algebra become h.w. states of the R algebra with helicity
 (+) (\ie\ annihilated by $G_0^+$). In addition, 
$\cU_{1/2}$ and $\cA_{-1/2}$ map the chiral primaries of
the NS algebra (\ie\ annihilated by $G^+_{-1/2}$) into
the set of R ground states, whereas
$\cU_{-1/2}$ and $\cA_{1/2}$ map the antichiral primaries
(\ie\ annihilated by $G^-_{-1/2}$) into the set of
R ground states. As a result, the spectral flows
\req{spfl} and  \req{ospfl}
 transform the NS singular vectors 
built on chiral and antichiral primaries into R singular vectors
built on the R ground states. 

\subsection{Spectrum of Singular Vectors}\lvm

We showed in section 2 that the NS singular vectors built on chiral
primaries are only of two types, $\ket{\chi_{NS}}_l^{(0)ch}$
and $\ket{\chi_{NS}}_{l-\half}^{(-1)ch}$, and similarly, the
NS singular vectors built on antichiral primaries come only
in two types $\ket{\chi_{NS}}_l^{(0)a}$ and 
$\ket{\chi_{NS}}_{l-\half}^{(1)a}$. In fact, $\ket{\chi_{NS}}_l^{(0)ch}$
and $\ket{\chi_{NS}}_l^{(0)a}$, on the one hand, and
$\ket{\chi_{NS}}_{l-\half}^{(-1)ch}$ and $\ket{\chi_{NS}}_{l-\half}^{(1)a}$,
on the other hand, are mirrored under the interchange
 $H_m \leftrightarrow -H_m$, $G_r^+ \leftrightarrow G_r^-$, because
they are the two possible untwistings of the \Gn -closed
topological singular vectors $\ket{\chi_T}^{(0)G}_l$ and 
$\ket{\chi_T}^{(1)G}_l$ respectively. 

{}From the results just discussed one deduces straightforwardly that 
the NS singular
vectors $\ket{\chi_{NS}}_l^{(0)a}$, $\ket{\chi_{NS}}_{l-\half}^{(1)a}$,
$\ket{\chi_{NS}}_l^{(0)ch}$ and 
$\ket{\chi_{NS}}_{l-\half}^{(-1)ch}$ are transformed, 
by the spectral flows \req{spfl} and \req{ospfl}, into R singular
vectors built on R ground states, in the following way:

\BE \cA_{1/2} {\ } \ket{\chi_{NS}}_l^{(0)a} = 
\cU_{1/2} {\ } \ket{\chi_{NS}}_l^{(0)ch}=\ket{\chi_R}_l^{(0)-} \EE

\BE \cA_{1/2} {\ } \ket{\chi_{NS}}_{l-\half}^{(1)a} = 
\cU_{1/2} {\ } \ket{\chi_{NS}}_{l-\half}^{(-1)ch}=\ket{\chi_R}_l^{(-1)-}\EE

\BE \cU_{-1/2} {\ } \ket{\chi_{NS}}_l^{(0)a} = 
\cA_{-1/2} {\ } \ket{\chi_{NS}}_l^{(0)ch}=\ket{\chi_R}_l^{(0)+} \EE

\BE \cU_{-1/2} {\ } \ket{\chi_{NS}}_{l-\half}^{(1)a} = 
\cA_{-1/2} {\ } \ket{\chi_{NS}}_{l-\half}^{(-1)ch}=\ket{\chi_R}_l^{(1)+}\EE

\noi
with $\ket{\chi_R}^{(0)+}$ and $\ket{\chi_R}^{(0)-}$ on the one hand, and
$\ket{\chi_R}^{(1)+}$ and $\ket{\chi_R}^{(-1)-}$ on the other hand, 
mirrored to each other.
Observe that the charged singular vectors built on the R ground
states come only in two types: $\ket{\chi_R}_l^{(1)+}$ and
$\ket{\chi_R}_l^{(-1)-}$, \ie\ the sign of the relative charge is
equal to the sign of the ``helicity".
The chiral and antichiral
NS Verma modules $V_{NS}^{ch}(\htop)$ and $V_{NS}^{a}(\htop)$
are transformed, in turn, in R Verma modules built on R ground states as 

\vskip .17in
${\ }{\ }{\ }$
${\ }{\ }{\ } \cU_{1/2}{\ } V_{NS}^{ch}(\htop)
 \rightarrow V_R(\htop-{\ctop\over6}){\ }$, 
${\ }{\ }{\ }\cU_{-1/2}{\ }V_{NS}^{a}(\htop)
 \rightarrow V_R(\htop+{\ctop\over6}){\ }$,\\
${\ }{\ }{\ }{\ }$
${\ }{\ }{\ }\cA_{1/2}{\ } V_{NS}^{a}(\htop)
 \rightarrow V_R(-\htop-{\ctop\over6}){\ }$,
${\ }{\ }{\ }\cA_{-1/2}{\ }V_{NS}^{ch}(\htop)
 \rightarrow V_R(-\htop+{\ctop\over6}){\ }$.

\vskip .17in

Hence, whereas the spectrum of conformal weights for all the R Verma
modules built on R ground states
is just given by $\Delta={\ctop\over24}$, the 
spectrum of U(1) charges for those which contain singular vectors is given
by: 
${\ }\htop_{r,s}^{(0)+} = \htop_{r,s}^{(0)} + {\ctop\over6}{\ }$
for singular vectors of type $\ket{\chi_R}_l^{(0)+}$,
${\ }\htop_{r,s}^{(1)+} = \htop_{r,s}^{(1)} + {\ctop\over6}{\ }$
for singular vectors of type $\ket{\chi_R}_l^{(1)+}$,
${\ }\htop_{r,s}^{(0)-} = - (\htop_{r,s}^{(0)} + {\ctop\over6}){\ }$
for singular vectors of type $\ket{\chi_R}_l^{(0)-}$
and ${\ }\htop_{r,s}^{(-1)-}=-(\htop_{r,s}^{(1)} + {\ctop\over6}){\ }$
for singular vectors of type $\ket{\chi_R}_l^{(-1)-}$.
Using the expressions for ${\ }\htop_{r,s}^{(0)}{\ }$
 and ${\ }\htop_{r,s}^{(1)}{\ }$
given by \req{h0rs} and \req{h1rs} one obtains:

\BE \htop_{r,s}^{(0)+} =  \htop_{r,s}^{(-1)-} =
-\half \L( \L({\ctop-3\over3} \R)r+s-1 \R)
 {\ } {\rm for} \qquad \ket{\chi_R}_l^{(0)+}{\ } 
  {\rm and}{\ }  \ket{\chi_R}_l^{(-1)-}{\ } 
 \label{h0+rs}, \EE

\BE \htop_{r,s}^{(0)-} =  \htop_{r,s}^{(1)+} =
\half \L( \L({\ctop-3\over3} \R)r+s-1 \R)
 {\ } {\rm for} \qquad \ket{\chi_R}_l^{(0)-}{\ } 
  {\rm and}{\ } \ket{\chi_R}_l^{(1)+}{\ }{\ }. 
 \label{h0-rs} \EE

\noi

We see that $\htop_{r,s}^{(0)+} =  \htop_{r,s}^{(-1)-} =
-\htop_{r,s}^{(0)-} =  -\htop_{r,s}^{(1)+}$. Therefore
the singular vectors of the types $\ket{\chi_R}_l^{(0)+}{\ }$
and $\ket{\chi_R}_l^{(-1)-}{\ }$
are together in the same Verma module $V_R(\htop)$ and at the same level,
and so are the mirrored singular vectors of the types
$\ket{\chi_R}_l^{(0)-}{\ }$ and $\ket{\chi_R}_l^{(1)+}{\ }$,
which belong to the mirror Verma module $V_R(-\htop)$. 
In other words, the 
singular vectors built on the R ground states come always
in sets of two mirrored pairs at the same level. Every pair consists of
one charged and one uncharged vector, with opposite helicities,
one pair belonging to the Verma module $V_R(\htop)$ and the
mirrored pair belonging to the mirror Verma module $V_R(-\htop)$.

This family structure is easy to see also
taking into account that the action of $G_0^+$ or $G_0^-$ on
any R singular vector built on the R ground states produces
another singular vector with different relative charge and different
helicity but with the same level and sitting in the same Verma
module. This resembles very much the family structure for the 
topological singular vectors 
built in chiral Verma modules, that we analyzed in section 2. However, 
there is a drastic difference here because in the latter case the
four members of the topological family, \ie\ 
$\ket{\chi_T}_l^{(0)G}$, $\ket{\chi_T}_l^{(0)Q}$, 
$\ket{\chi_T}_l^{(1)G}$ and $\ket{\chi_T}_l^{(-1)Q}$,
are completely different from each other, while in this case the four
 members are two by two mirror symmetric under the interchange 
$H_m \leftrightarrow -H_m{\ }$, $G_r^+ \leftrightarrow G_r^-$.

\subsection{Subsingular Vectors}\lvm
Now let us compare the spectra we have found, eqns. \req{h0+rs}
and \req{h0-rs}, with the roots of the determinant 
formula for the R algebra \cite{BFK}, \cite{KaMa3}, specialized 
to the case $\Delta = {\ctop\over 24}$. 
The roots of the determinant formula for the R algebra are given, 
in ref. \cite{BFK}, by the vanishing quadratic surface $f_{r,s}^P=0$
and the vanishing plane $g_k^P=0$, where

\BE f_{r,s}^P = 2\L({\ctop-3\over3}\R)(\Delta-{\ctop\over24}) - \htop^2
+{1\over4}\L(\L({\ctop-3\over3}\R) r+s \R)^2 \qquad r\in\oZ^+\,,\,\,
  s\in2\oZ^+ \label{frsP} \EE

\noi
and 
       
\BE   g_k^P = 
  2 \Delta-2k\htop + \L({\ctop-3\over3}\R)(k^2-{1\over4})
 - {1\over4} \qquad k\in\oZ+\half \,. \EE

\noi
For $\Delta={\ctop\over24}$ they result in the following solutions

\BE  \htop_{r,s}^{BFK} = \pm {\ } \half \L(
      \L({\ctop-3\over3}\R) r + s \R)  \label{hrsB} \EE

\noi
and

\BE  \htop_k^{BFK} = {\ } \half 
      \L({\ctop-3\over3}\R) k  \label{hkB} \EE

\noi
where the superscript BFK indicates that  these are U(1) charges in the 
notation of ref. \cite{BFK}. Therefore we have to add $(\pm \half)$
to these expressions to translate them into our notation. Namely,

\BE  \htop_{r,s}^{BFK} + \half=
 \ccases{\half \L( \L({\ctop-3\over3}\R) r + s + 1 \R)}
{-\half \L( \L({\ctop-3\over3}\R) r + s - 1 \R)}   \EE

\BE  \htop_{r,s}^{BFK} - \half=
 \ccases{\half \L( \L({\ctop-3\over3}\R) r + s - 1 \R)}
{-\half \L( \L({\ctop-3\over3}\R) r + s + 1 \R)}   \EE

\noi
and

\BE  \htop_k^{BFK} + \half {\ } = {\ }
 \half \L( \L({\ctop-3\over3}\R) k + 1 \R) 
\EE

\BE  \htop_k^{BFK} - \half {\ } = {\ }
 \half \L( \L({\ctop-3\over3}\R) k - 1 \R) 
\EE

Comparing these expressions with the spectra given by \req{h0+rs} and
\req{h0-rs}, we see that the situation
is the same as for the NS algebra for Verma modules built
on chiral primaries. Namely, half of the zeroes of 
$f_{r,s}^P=0$, for every pair $(r,s)$, 
correspond to uncharged singular vectors and the other
half correspond to charged singular vectors. However, since now
charged and uncharged singular vectors (with different helicities) 
share the same spectra, this statement is rather ambiguous unless we
specify the helicities. Let us make the choice that adding $\half$
(or $-\half$) to the BFK spectra \req{hrsB} and \req{hkB} one gets
the spectra corresponding to the helicity $+$ (or $-$) singular 
vectors. Then one finds the following identifications. The upper solution
of $\htop_{r,s}^{BFK} + \half$ corresponds to the charged
singular vectors $\ket{\chi_R}_l^{(1)+}$  at level ${\ }r(s+2) \over2{\ }$,
 whereas the lower solution corresponds to the uncharged
singular vectors $\ket{\chi_R}_l^{(0)+}$ at level $rs\over2{\ }$.
The upper solution of $\htop_{r,s}^{BFK} - \half$  corresponds to
the uncharged singular vectors
 $\ket{\chi_R}_l^{(0)-}$ at level $rs\over2{\ }$,
while the lower solution corresponds to the charged
singular vectors $\ket{\chi_R}_l^{(-1)-}$
 at level $r(s+2)\over2{\ }$. The solutions $\htop_k^{BFK}+\half$
 and $\htop_k^{BFK}-\half$ correspond to the charged
singular vectors $\ket{\chi_R}_l^{(1)+}$ at level $k$, and 
$\ket{\chi_R}_l^{(-1)-}$ at level $(-k)$, respectively.

Therefore, in analogous way as happens for the NS 
algebra, the zeroes of the vanishing plane $g_k^P=0$ give the
solutions  $\htop_{r,2}^{(1)+}$ and 
$\htop_{r,2}^{(-1)-}{\ }$ (\ie\ $\pm \htop_{r,2}^{(1)+}$),
while half of the zeroes of $f_{r,s}^P=0$, 
for every pair $(r,s)$, give the solutions 
$\htop_{r,s>2}^{(1)+}$ and 
$\htop_{r,s>2}^{(-1)-}{\ }$ (\ie\ $\pm \htop_{r,s>2}^{(1)+}$),
corresponding to the charged singular vectors of types 
$\ket{\chi_R}^{(1)+}_l$ and $\ket{\chi_R}^{(-1)-}_l$.
As a result, the complete R Verma modules, with $\D={\ctop\over 24}$ 
and $\htop=\pm \htop_{r,s>2}^{(1)+}$, 
contain charged subsingular vectors of types
$\ket{\chi_R}^{(1)+}_l$ and $\ket{\chi_R}^{(-1)-}_l$, respectively.

\section{N=2 Chiral Determinant Formulae}\lvm

In sections 3 and 4 we have derived conjectures for the roots of the
N=2 chiral determinants, corresponding to chiral topological
Verma modules, chiral and antichiral NS Verma modules, and R Verma
modules built on the R ground states. Using these conjectures, 
together with some computation and some consistency
checks, described below, we have obtained the following expressions 
for the N=2 chiral determinant formulae.

\vskip .18in

\noi
{\it Topological algebra}
\vskip .08in

The chiral topological Verma modules satisfy 
$\D=0$. The chiral topological determinant formula, in terms of the U(1) 
charges, is given by the expression:

\BE
det(\cM^T_l) = {\rm cst.} \prod_{2\leq rs \leq 2l}(\htop-\htop_{r,s}^{(0)})^
{2P_r^T(l-{rs\over2})}{\ \ }(\htop-\htop_{r,s}^{(1)})^
{2P_r^T(l-{rs\over2})}{\ \ \ }r\in\oZ^+,{\ }s\in2\oZ^+ \,, \label{Tdet}
\EE

\noi
with the roots $\htop_{r,s}^{(0)}$ and $\htop_{r,s}^{(1)}$, given by
eqns. \req{h0rs} and \req{h1rs}, satisfying
$\htop_{r,s}^{(1)}=-\htop_{r,s}^{(0)}-\ctop/3$. The factors 2 in
the exponents show 
that the topological singular vectors in chiral Verma modules come 
two by two, one charged and one uncharged, at the same level in the
same Verma module. Namely,
the singular vectors of types $\ket{\chi_T}_l^{(0)G}$ and
$\ket{\chi_T}_l^{(-1)Q}$ are always together, and the same happens with 
the singular vectors of types $\ket{\chi_T}_l^{(1)G}$ and
$\ket{\chi_T}_l^{(0)Q}$.

\noi
The partitions $P_r^T$ are defined by

\BE
\sum_N P_r^T(N)x^N={1\over 1+x^r}{\ }\sum_n P^T(n)x^n=
{1\over 1+x^r}{\ } \prod_{0<l\in {\bf Z},{\ }0<m\in {\bf Z}} {(1+x^l)^2 \over 
(1-x^m)^2}\,.
\label{PT}
\EE

\vskip .18in
\noi
\newpage

{\it Antiperiodic NS algebra}

\vskip .08in 

The chiral and antichiral NS Verma
modules satisfy $\D=\pm\htop/2$, respectively. The chiral and antichiral 
NS determinant formulae are  given by a single expression in terms
of the conformal weights: 

\BE
det(\cM_l^{NS-ch})=det(\cM_l^{NS-a})= {\rm cst.}
\prod_{2\leq rs \leq 2l}(\D-\D_{r,s}^{(0)})^
{P_{r+{1\over2}}^{NS}(l-{rs\over2})}
\prod_{0<k={rs-1\over2}\leq l}(\D-\D_{r,s}^{(1)})^
{P_{r-{1\over2}}^{NS}(l-k)} ,
\label{NSdet1}
\EE

\noi
and by two different expressions in terms of the U(1) charges:

\BE
det(\cM_l^{NS-ch}) = {\rm cst.} \prod_{2\leq rs \leq 2l}(\htop+\htop_{r,s}^{(0)})^
{P_{r+{1\over2}}^{NS}(l-{rs\over2})}
\prod_{0<k={rs-1\over2}\leq l}(\htop+\htop_{r,s}^{(1)})^
{P_{r-{1\over2}}^{NS}(l-k)}{\ \ \ }r\in\oZ^+,{\ }s\in2\oZ^+ \,,
\label{NSdet2} \EE

\BE
det(\cM_l^{NS-a}) = {\rm cst.} \prod_{2\leq rs \leq 2l}(\htop-\htop_{r,s}^{(0)})^
{P_{r+{1\over2}}^{NS}(l-{rs\over2})}
\prod_{0<k={rs-1\over2}\leq l}(\htop-\htop_{r,s}^{(1)})^
{P_{r-{1\over2}}^{NS}(l-k)}{\ \ \ }r\in\oZ^+,{\ }s\in2\oZ^+ \,,
\label{NSdet3} \EE

\noi
where $\D_{r,s}^{(0)}=-\htop_{r,s}^{(0)}/2$,
$\D_{r,s}^{(1)}=-\htop_{r,s}^{(1)}/2$, and  ${\ }\htop_{r,s}^{(0)}$ and
$\htop_{r,s}^{(1)}$ are given by eqns. \req{h0rs} and \req{h1rs}, like
for the Topological algebra.

\noi
The partitions $P_r^{NS}$ are defined by

\BE
\sum_N P_r^{NS}(N)x^N={1\over 1+x^r}{\ }\sum_n P^{NS}(n)x^n=
{1\over 1+x^r}{\ }\prod_{0<k\in {\bf Z}+{1\over2},{\ }0<m\in {\bf Z}}{(1+x^k)^2 \over
(1-x^m)^2} .
\EE

\vskip .18in
\noi
{\it Periodic R algebra}

\vskip .08in 
The R Verma modules built on R
ground states satisfy $\D={\ctop \over 24}$. The corresponding R
determinant formula,
in terms of the U(1) charges, is given by the expression:

\BE
det(\cM^R_l) = {\rm cst.} \prod_{2\leq rs \leq 2l}(\htop-\htop_{r,s}^{(0)+})^
{2P_r^R(l-{rs\over2})}{\ }(\htop-\htop_{r,s}^{(1)+})^
{2P_r^R(l-{rs\over2})}{\ \ \ }r\in\oZ^+,{\ }s\in2\oZ^+ \,,
\label{Rdet}
\EE

\noi
with the roots $\htop_{r,s}^{(0)+}$ and $\htop_{r,s}^{(1)+}$, given by
eqns. \req{h0+rs} and \req{h0-rs}, satisfying
$\htop_{r,s}^{(1)+}=-\htop_{r,s}^{(0)+}$.
The factors 2 in the exponents show that the
R singular vectors built on R ground states come 
two by two, one charged and one uncharged, at the same level in the
same Verma module. That is,
the singular vectors of types $\ket{\chi_R}_l^{(0)+}$ and
$\ket{\chi_R}_l^{(-1)-}$ on the one hand, and 
the singular vectors of types $\ket{\chi_R}_l^{(1)+}$ and
$\ket{\chi_R}_l^{(0)-}$, on the other hand, are always together.
The partitions $P_r^R$ coincide exactly with
the partitions corresponding to the Topological algebra, \ie\
$P_r^R=P_r^T$, defined in equation \req{PT}. Therefore the
exponents are identical for the chiral topological determinants 
\req{Tdet} as for the R determinants \req{Rdet}. 

This fact is easy to understand taking into account 
that the deformation of the chiral and antichiral NS Verma modules 
into chiral topological Verma modules, under the twists $T_{W2}$ \req{twb}
and $T_{W1}$ \req{twa} respectively, follows the same pattern as the
deformation  of the NS Verma modules into R Verma modules under the
spectral flows. In particular,
the reorganization of states for the corresponding Verma modules,
at every 
level, satisfies $l^T=l^R=l^{NS} + \half |q^{NS}|$, so that the number
of states in the topological Verma modules is equal to the number of
states in the R Verma modules level by level.
\vskip .20in
\noi
{\it Consistency Checks}

\vskip .08in 
We have performed consistency checks on the N=2 chiral determinant
formulae\footnote{suggested to us by Adrian Kent}. They are based on 
the fact that, for the NS algebra and for the R algebra,
the terms of the chiral determinants with highest power of $\ctop$ are 
the diagonal terms (similar statements cannot be made for $\D$ or $\htop$).

Let us start with the R algebra, which is the easiest case. A
simple exercise shows that the power of $\ctop$ in the diagonal
of the level $l$ matrix, for Verma modules built on R ground
states, is given by the coefficients of
the generating function

\BE
\prod_{0<m\in {\bf Z}}{(1+x^m)^2\over(1-x^m)^2} {\ }
\sum_{0<n \in {\bf Z}}{4x^n\over1-x^{2n}}{\ \ \ }  \,.
\label{cc1}
\EE

\noi
These coefficients are to be compared with the highest power of  
 $\ctop$ predicted by the determinant formula \req{Rdet}, that is

\BE
4\sum_{2\leq rs\leq2l}P^R_r(l-{rs\over2}) \,,
\label{cc2}
\EE

\noi
where one takes into account that $\htop_{r,s}^{(0)+}$ and
$\htop_{r,s}^{(1)+}$ 
contribute with one $\ctop$ each. One obtains that the coefficients
of the generating function \req{cc1} can be expressed exactly as
\req{cc2}, after a rather straightforward manipulation, using the
definitions in \req{PT}.

One finds analogous results, although more laborious, for the
NS algebra. The power of $\ctop$ in the diagonal of the level
$l$ matrix for chiral Verma modules is given by the coefficients
of the generating function

\BE
{1\over(1+x^{1/2})}
\prod_{0<n\in {\bf Z}}{1\over(1-x^n)^2}
\prod_{0<k\in {\bf Z}+1/2}(1+x^k)^2
\L\{\sum_{0<l\in {\bf Z}}{2 x^l\over1-x^l}-{x\over1-x}+
\sum_{{1\over2}<k\in {\bf Z}+1/2}{2 x^k\over1+x^k}\R\} \,.
\label{cc3}
\EE

\noi
These coefficients coincide with the highest power of $\ctop$ predicted 
by the determinant formulae \req{NSdet1}-\req{NSdet3}, that is

\BE
\sum_{2\leq rs\leq2l}P^{NS}_{r+1/2}(l-{rs\over2})+
\sum_{0<k={rs-1\over2}\leq l,\,  r>1}P^{NS}_{r-1/2}(l-k) \,,
\label{cc4}
\EE

\noi
where one takes into account that $\htop_{r,s}^{(0)}$ contributes with
one $\ctop$ 
each and $\htop_{r,s}^{(1)}$ contributes with one $\ctop$ 
except for $r=1$.

For the Topological algebra there is no need to make an independent
consistency check. The reason is that, on the one hand,
the roots of the chiral topological determinants are equal to the roots
of the antichiral NS determinants, as we showed in section 2, 
whereas, on the other hand,
the exponents of the topological determinants are equal to
the exponents of the R determinants, as we just discussed.

\section{Conclusions and Final Remarks}\lvm

{\it First}, we have analyzed 
in much detail the relation between the singular 
vectors of the Topological algebra and the singular vectors of the 
NS algebra, showing the direct relation between their 
corresponding spectra. Then we have deduced, using the family structure of 
the topological singular vectors, 
that charged and uncharged NS singular vectors in chiral 
NS Verma modules come in pairs, although in different Verma modules,
with a precise relation between them: 
$V^{ch}_{NS}(\htop) \leftrightarrow V^{ch}_{NS}(-\htop+\ctop/3)$ and
$V^{a}_{NS}(\htop) \leftrightarrow V^{a}_{NS}(-\htop-\ctop/3)$ for
chiral and antichiral NS Verma modules, respectively. 
This result contrasts drastically the case
of complete Verma modules, where the charged singular vectors,
which correspond to a one-parameter family of roots of the NS
determinant formula, are much less numerous than the uncharged 
singular vectors, 
which correspond to a two-parameter family of roots of the NS
determinant formula.
In addition, we have shown that the charged NS singular vectors
 built on chiral primaries have always relative charge $q=-1$,
while those built on antichiral primaries have always relative
charge $q=1$. These vectors are mirrored to each other under the 
interchange $H_m \leftrightarrow -H_m,{\ \ }G_r^+ \leftrightarrow G_r^-$
because they are the two possible untwistings of the same topological 
singular vectors of type $\ket{\chi_T}^{(1)G}$.

\vskip .17in

{\it Second}, using the result 
that charged and uncharged singular vectors in chiral
Verma modules come in pairs, we have derived a conjecture 
for the roots of the chiral
determinant formulae for the Topological algebra and the NS algebra. 
For this purpose we have also made the ansatz that the roots of the
chiral determinant formulae coincide with the roots of the 
determinant formulae specialized to the values of 
(or relations between) the
conformal weights and U(1) charges which occur in chiral Verma modules 
($\D=0$ for chiral topological Verma modules and $\D=\pm\htop/2$ 
for chiral and antichiral NS Verma modules, respectively).
Our results imply that both uncharged and charged singular vectors,
correspond to two-parameter families of roots of the
chiral determinant formulae, denoted as $\htop^{(0)}_{r,s}$ 
eq. \req{h0rs}, and $\htop^{(1)}_{r,s}$ eq. \req{h1rs}, respectively, 
and agree with all known data (spectrum
of topological and NS singular vectors in chiral Verma modules from
level 1/2 to level 4). 

Our results also imply the existence of subsingular
vectors in the complete Verma modules. These are the singular vectors
which are singular only in the chiral Verma modules, becoming
non-highest weight null vectors in the complete Verma modules, which
are not descendants of any singular vectors. We have found
that there are {\it charged subsingular} vectors, at levels $rs-1\over2$, 
in the complete NS Verma modules with $\D= \mp \htop/2$ for
$\htop = \pm \htop_{r,s>2}^{(1)}$, which
contain uncharged singular vectors at levels $r(s-2)\over2$. 
Once the chirality
is imposed, what amounts to ``divide" the complete NS Verma
modules by the singular vectors 
${\ }G_{-1/2}^-\ket{\D,\htop}$, with $\D=-\htop/2$, or
${\ }G_{-1/2}^+\ket{\D,\htop}$, with $\D=\htop/2{\ }$,
the uncharged singular vectors vanish whereas the charged subsingular
vectors become singular in the chiral Verma modules.
Observe the asymmetry of r\^oles between ``charged" subsingular vectors,
which become singular, and ``uncharged" singular vectors, which vanish,
once the chirality is switched on. This asymmetry does not exist in
the Topological algebra where there are both {\it charged and
uncharged subsingular} vectors, and both uncharged and charged singular
vectors which vanish when the chirality is imposed and the
subsingular vectors become singular. The complete symmetry, in the 
case of the Topological algebra, is reflected in the fact that the
topological subsingular vectors are located in complete
topological Verma modules with $\D=0$,  $\,V(\ket{0,\,\htop}^G)$ and
$V(\ket{0,\,\htop}^Q)$, with $\htop=\htop_{r,s>2}^{(1)}$ and  
$\htop=\htop_{r,s>2}^{(0)}$, respectively. Under the untwistings only
the topological h.w. states (primaries and singular vectors) which
are \Gn-closed remain h.w. states of the NS algebra. For this reason,
the symmetry between charged and uncharged topological subsingular
vectors is broken in the NS algebra.

\vskip .17in

{\it Third}, using the spectral flows between the NS algebra and
the R algebra we have translated all the results which apply
to chiral NS Verma modules, into results which apply to
R Verma modules built on Ramond ground states. In particular we
have obtained  the 
corresponding conjecture for the roots of the determinant formula 
corresponding to this type of R Verma modules (for which 
$\D=\ctop/24$). The situation with respect to subsingular vectors
for the R algebra is analogous to the situation for the NS algebra.
Namely, we have found {\it charged subsingular} vectors in complete R
Verma modules with $\D=\ctop/24$ and 
$\htop = \pm \htop_{r,s>2}^{(1)+}$, given by eq. \req{h0-rs}.

\vskip .17in

{\it Finally}, using some 
computer exploration as well as the conjectures for
the corresponding roots, we have written down expressions for the
chiral determinant formulae, \ie\ for the chiral topological Verma
modules, for the chiral and antichiral NS Verma modules and for the 
R Verma modules built on the Ramond ground states. We have provided
consistency checks for these formulae.
\vskip .17in

It is already remarkable the fact that the roots of the chiral
determinants coincide (as far as we can tell) with the roots
of the determinants for the specific relations between
$\D$ and $\htop$. We have not found any proof for this ansatz.
Also remarkable is the fact that, in the chiral Verma modules, 
only half of the zeroes of the quadratic vanishing surfaces
$f_{r,s}=0$, for every pair $(r,s)$,
correspond to uncharged singular vectors, while the other half
of the solutions 
correspond to charged singular vectors, in contrast with
 the case of complete Verma modules for which all the
solutions to $f_{r,s}=0$ correspond to uncharged singular vectors.
Although this remarkable behaviour of the roots of the N=2 determinants
is due to the existence of subsingular vectors, 
it seems to us that there
 exists a ``singular vector conservation law"
or ``singular vector transmutation" when switching
on and off chirality on the primary states, with charged
and uncharged singular vectors replacing each other.

\vskip .17in

In the Appendix we have analyzed thoroughly the NS singular 
(and/or subsingular) vectors 
$\ket{\chi_{NS}}_1^{(0)}$, $\ket{\chi_{NS}}_{3\over2}^{(1)}$ and 
$\ket{\chi_{NS}}_{3\over2}^{(-1)}$. We have written down the h.w. equations, 
with their solutions, for the cases of the primary states being non-chiral, 
chiral and antichiral. We also have investigated the behavior of the
subsingular vectors, and we have attempted to shed some light on the issue
of the ``singular vector transmutation". 

\vskip .17in

As a final remark, we stress the fact that we have discovered 
subsingular vectors in the N=2 Superconformal algebras,
their very existence was completely unknown,
and we have written down examples. Subsingular vectors do not exist
in the Virasoro algebra neither in the sl(2) algebra. The 
subsingular vectors that we have found are those 
which become singular in the 
chiral Verma modules. The issue whether these are the only
subsingular vectors in the N=2 Superconformal algebras
is currently under investigation.

\vskip .45in

\centerline{\bf Acknowledgements}

We thank very much A. Kent for many explanations and suggestions  
about determinant formulae and related issues. We also thank very
much A.N. Schellekens
for many illuminating discussions; his insight was essential
for the discovery of subsingular vectors.
We also thank the Tata Institute of Fundamental Research (Bombay)
for hospitality,
where an important part of this paper was worked out. In particular
we thank S. Mukhi and A.K. Raina for useful comments to this work.
We are indebted to J. Gaite, A.N. Schellekens, Ch. Schweigert, 
and especially J. Vermaseren for help with the computer. 
The chiral determinants have been computed using FORM. Finally, 
we are very grateful to the referee of Nuclear Physics for reading 
our work with so much interest and giving us many useful suggestions  
to improve our paper (see the note below).

\subsection*{Important Note}

This is an extended and improved version of the work presented last year
in hep-th/9602166. Following the suggestions of the referee of
Nuclear Physics B: {\,} i) we have derived conjectures for the whole N=2
chiral determinant formulae, not only for the roots,
and {\,} ii) we have given appropriate emphasis to the importance
of the discovery of subsingular vectors for the N=2 Superconformal
algebras. Between the first version and the present version 
we have published some examples of subsingular vectors for
the Topological algebra in ref \cite{BJI6}. 
To our knowledge no other authors have given
examples of subsingular vectors for the N=2 Superconformal algebras.
Recently a paper has appeared by A.M. Semikhatov and I.Yu. Tipunin
(hep-th/9704111), dealing with N=2 subsingular vectors, in which
there are many claims for which no proofs are given. Furthermore
we disagree with several statements. 

\setcounter{equation}{0}
\def\theequation{A.\arabic{equation}}

\subsection*{Appendix}

In what follows we give a very detailed example of the differences
between singular vectors in chiral NS Verma modules
and singular vectors in complete NS Verma modules.
We present the highest weight (h.w.) conditions, with
their solutions, which define the singular vectors
$\ket{\chi_{NS}}_{3\over2}^{(1)}$,{\ } $\ket{\chi_{NS}}_{3\over2}^{(-1)}$ and
{\ }$\ket{\chi_{NS}}_1^{(0)}$, for the cases when the primary state
$\ket{\D,\htop}$ is: $\,$ i) non-chiral, indicating the results for
$\D=\pm\htop/2$, $\,$ ii) chiral, \ie\ which satisfy 
$G^+_{-1/2}\ket{\D,\htop}=0$, $\D=\htop/2$, and $\,$ iii) antichiral, \ie\
which satisfy $G^-_{-1/2}\ket{\D,\htop}=0$, $\D=-\htop/2$. 
The subsingular vectors are the states
which are singular in the chiral or antichiral Verma modules, but not in
the complete Verma modules. We show that these states, which correspond to
the roots $\pm \hat\htop_{r,s}$ in \req{hh0rs}, or equivalently 
$\pm \htop_{r,s>2}^{(1)}$ in \req{h1rs}, are non-h.w. null 
vectors in the complete Verma modules, not descendants of any
singular vectors.

\vskip 0.20in
\noi
{\it The singular/subsingular vectors} $\ket{\chi_{NS}}_{3\over2}^{(1)}$

\vskip .08in
\noi
The general form of the charge $q=1$ NS singular vectors at level 
${3\over2}$ is

\BE 
\ket{\chi_{NS}}_{3\over 2}^{(1)}=(\alpha L_{-1}G_{-1/2}^++
\beta H_{-1}G_{-1/2}^++\gamma G_{-3/2}^+)\ket{\Delta, \htop}\,.
\EE

\noi
The h.w. conditions
 $L_{m>0}\kc=H_{m>0}\kc=G_{r\geq\half}^+ \kc=
G_{r\geq\half}^- \kc=0$, which
 determine the coefficients ${\ }\alpha,{\ } \beta,{\ } \gamma, {\ }$ 
as well as the conformal weight $\Delta$  and the U(1) charge $\htop$
of the primary state, result as follows.
For $\ket{\Delta, \htop}$ non-chiral one obtains the 
equations

\begin{eqnarray}
\alpha(1+2\Delta)+\beta(1+\htop)+2\gamma &=& 0\nonumber\\
\alpha(1+\htop)+\beta{\ }{\ctop\over 3}+\gamma &=& 0\nonumber\\
(2\alpha+\beta)(2\Delta-\htop)+
\gamma(2\Delta-3\htop+2{\ctop\over3}) &=& 0\nonumber\\
\beta(2\Delta-\htop)-2\gamma &=& 0\nonumber\\
\alpha(2\Delta-\htop)+2\gamma &=& 0\nonumber\\
\alpha+\beta &=& 0\,.\label{eq3n}
\end{eqnarray}

\noi
We see that, for $\Delta\neq{\htop\over2}$, ${\ }\gamma$
must necessarily be different from zero 
(if $\gamma=0$ the whole vector vanishes). Hence
we can choose $\gamma=1$.
Solving for the other coefficients one obtains the solution, for 
$\Delta\neq{\htop\over2}$

\BE
\alpha={-2\over2\Delta-\htop}{\ \ }, \qquad \beta={2\over2\Delta-\htop}
\label{sol3n}\EE

\noi
with $\Delta-{3\over2}\htop+{\ctop-3\over3}=0$. This solution is given by 
the vanishing plane $g_{3/2}=0$, as one can check in eq. \req{gk}.

For the case $\Delta={\htop\over2}$ the solution is $\gamma=0$, $~\beta=-\alpha$ 
and $\htop={\ctop-3\over3}$. It is also given by the vanishing plane 
$g_{3/2}=0$. If we now specialize the general solution
\req{sol3n} to the case 
$\Delta=-{\htop\over2}$ we find

\begin{eqnarray}
 \alpha={6\over\ctop-3}{\ },\qquad\beta={6\over3-\ctop}{\ },\qquad
\htop={\ctop-3\over6}\,{\ }. \label{sol3na}
\end{eqnarray}

{\it For} $\ket{\Delta, \htop}$ {\it chiral}
{\ }the h.w. conditions only give the equation
 $\gamma=0$. Therefore $\ket{\chi_{NS}}_{3\over 2}^{(1)ch}$ vanishes
while $\ket{\chi_{NS}}_{3\over 2}^{(1)}$, for $\Delta={\htop\over2}$,
is a singular vector with $\g=0,{\ }\b=-\a$, as we have just
 shown. This is a particular case
 of the general result that the charged singular vectors
 built on chiral primaries have always relative charge $q=-1$,
whereas those built on antichiral primaries have always $q=1$.

\vskip 0.15in

{\it For} $\ket{\Delta, \htop}$
{\it antichiral}
one gets the equations

\begin{eqnarray}
     \alpha(1-\htop)+\beta(1+\htop)+2\gamma &=& 0\nonumber\\
     \alpha(1+\htop)+\beta{\ }{\ctop\over 3}+\gamma &=& 0\nonumber\\
 (2\alpha+\beta)\htop+\gamma(2\htop-{\ctop\over 3}) &=& 0\nonumber\\
  \b{\ }\htop+\gamma &=& 0\nonumber\\
\alpha(1-\htop)+\beta+\gamma &=& 0\,.\label{eq3a}
\end{eqnarray}

\noi
Comparing these with equations \req{eq3n}, setting
 $\Delta=-{\htop\over2}$,
we see that here there is one equation less and the first four
equations coincide. The last equation here and the two last 
equations in \req{eq3n} are different though. These equations
 correspond to the h.w. condition 
 $G_\half^-\ket{\chi_{NS}}_{3\over2}^{(1)}=0$.
As before $\gamma\neq 0$ necessarily, thus we set $\gamma=1$.
Solving for the other coefficients and for $\htop$ one 
obtains two solutions:

\BE
\a=\ccases{6\over\ctop-3}{\ctop-3\over6}\!,
\qquad\b=\ccases{6\over3-\ctop}{-1}\!,
\qquad\htop=\ccases{\ctop-3\over6}{1}\!. \label{sol3a}
\EE

\noi
The solution $\htop={\ctop-3\over 6}$ is the solution given by 
the vanishing plane $g_{3/2}=0$, and therefore the only 
solution  for $\Delta =-{\htop\over2},{\ }\ket{\Delta,\htop}$
non-chiral, as we have shown in eq. \req{sol3na}. It corresponds to
$\htop_{2,2}^{(1)}$ in eq. \req{h1rs}. The solution
 $\htop=1$ corresponds to ${\ }\hat\htop_{1,2}{\ }$ in eq. \req{hh0rs},
and to $~\htop^{(1)}_{1,4}~$ in eq. \req{h1rs},
given by the vanishing quadratic surface $f_{1,2}=0$.
We see therefore that the charged subsingular vector (in the complete
Verma module) corresponds to
$\htop_{r,s>2}^{(1)}$ whereas the solution $\htop_{r,2}^{(1)}$,
given by the vanishing plane, corresponds to a singular vector which 
is also singular in the complete Verma module.

\vskip .2in
\noi
{\it The singular/subsingular vectors} $\ket{\chi_{NS}}_{3\over2}^{(-1)}$

\vskip .08in
\noi
The general form of the charge $q=-1$ NS singular vectors at 
level ${3\over2}$ is

\BE
\ket{\chi_{NS}}_{3\over2}^{(-1)}=(\alpha L_{-1}G_{-1/2}^-+
\beta H_{-1}G_{-1/2}^-+\gamma G_{-3/2}^-) \ket{\Delta, \htop}\,.
\EE

\noi
Since this case is very similar to the previous one we will consider
only the chiral representations.
The h.w. conditions result in $\gamma=0$ for $\ket{\Delta, \htop}$ 
antichiral (therefore $\ket{\chi_{NS}}_{3\over2}^{(-1)a}$ vanishes),
 while for $\ket{\Delta, \htop}$ chiral one gets the equations

\begin{eqnarray}
    \alpha(1+\htop)+\beta(\htop-1)+2\gamma &=& 0\nonumber\\
    \alpha(\htop-1)+\beta{\ }{\ctop\over 3}-\gamma &=& 0\nonumber\\
(2\alpha-\beta)\htop+\gamma(2\htop+{\ctop\over 3}) &=& 0\nonumber\\   
\beta{\ }\htop+\gamma &=& 0\nonumber\\
 \alpha(\htop+1)-\beta+\gamma &=& 0\,,
\end{eqnarray}

\noi
where, as before, we can set $\gamma=1$. The other coefficients and the U(1) 
charge $\htop$ read

\BE
\a=\ccases{6\over\ctop-3}{\ctop-3\over6}\!,
\qquad\b=\ccases{6\over\ctop-3}{1}\!,
\qquad\htop=\ccases{3-\ctop\over6}{-1}\!.\label{sol3ch} \EE

\noi
These solutions correspond to
$(-\htop^{(1)}_{2,2})$ and $(-\htop^{(1)}_{1,4})$ respectively.
The first solution is given by the vanishing plane 
$g_{-3/2}=0$, and the second solution, given by the vanishing
quadratic surface $f_{1,2}=0$, corresponds to a subsingular
vector in the complete Verma module, as in the previous case.

We see that $\ket{\chi_{NS}}_{3\over2}^{(1)a}$ and 
$\ket{\chi_{NS}}_{3\over2}^{(-1)ch}$ are mirrored under the interchange 
$H_m\leftrightarrow -H_m,{\ \ }G_r^+\leftrightarrow G_r^-$, reflecting the 
fact that they are the two different untwistings of the same topological 
singular vectors $\ket{\chi_T}_2^{(1)G}$, given by

\BE
\ket{\chi_T}_2^{(1)G}=(\cG_{-2}+\a\cL_{-1}\cG_{-1}+\b \cH_{-1}\cG_{-1})
\ket\phi_{\htop}
\EE

\noi
with

\BE
\a=\ccases{6\over\ctop-3}{\ctop-3\over6}\!,
\qquad\b=\ccases{6\over3-\ctop}{-1}\!,
\qquad\htop=\ccases{\ctop-3\over6}{1}\!. \EE

\noi
as the reader can verify using the twists $T_{W1}$ \req{twa} and
$T_{W2}$ \req{twb}.

\vskip .2in
\noi
{\it The singular vectors} $\ket{\chi_{NS}}_1^{(0)}$

\vskip .08in
\noi
The general form of the uncharged NS singular vectors at 
level 1 is given by 

\BE
\ket{\chi_{NS}}_1^{(0)}=(\alpha L_{-1}+\beta H_{-1}+
\gamma G_{-1/2}^+G_{-1/2}^-)
\ket{\Delta, \htop}\,.
\EE

\noi
The h.w. conditions result as follows. For  
$\ket{\Delta, \htop}$ non-chiral one obtains the equations

\begin{eqnarray}
2 \,\a{\ }\Delta+\b{\ }\htop+\gamma(2\Delta+\htop) &=& 0\nonumber\\
\a{\ } \htop+\b{\ } {\ctop\over3}+\gamma(2\Delta+\htop) &=& 0\nonumber\\
\alpha-\beta-\gamma(2\Delta+\htop) &=& 0\nonumber\\
\alpha+\beta+\gamma(2\Delta-\htop+2) &=& 0\,.\label{1nch}
\end{eqnarray}

\noi
As before we can set $\gamma=1$ and we get

\BE
\alpha=\htop-1, \qquad \beta=-(2\Delta+1)\,,
\EE

\noi
with $\htop^2-2\Delta{\ctop-3\over3}-{\ctop\over3}=0$. This
solution corresponds to 
the quadratic vanishing surface $f_{12}=0$ in \req{frs}. It has been
given before in refs. \cite{Kir2} and \cite{YuZh}.

\vskip 0.16in

For $\Delta={\htop\over 2}$ the solutions are

\begin{eqnarray}
\alpha=\ccases{\ctop-3\over3}{-2}\!,{\ \ }
\hat\alpha=\ccases{\ctop+3\over3}{0}\!,{\ \ }
\beta=\ccases{-{\ctop+3\over3}}{0}\!,{\ \ }\htop=\ccases{\ctop\over3}{-1}\!,
\label{eq1nch} \end{eqnarray}

\noi
 $\a$ transforming into $\hat\a$ if we commute the term 
$G_{-\half}^+G_{-\half}^-\rightarrow G_{-\half}^-G_{-\half}^+$. These 
solutions correspond to $(-\htop_{1,2})$ and $(-\hat\htop_{1,2})$ in
eqns. \req{hrs} and \req{hh0rs}, respectively.

\vskip 0.16in

For $\Delta=-{\htop\over 2}$ the solutions are 

\begin{eqnarray}
\alpha=\ccases{-{\ctop+3\over3}}{0}\!,{\ \ }
\beta=\ccases{-{\ctop+3\over3}}{0}\!,{\ \ }\htop=\ccases{-{\ctop\over3}}{1}\!.
\label{eq1na} \end{eqnarray}

\noi
These solutions correspond to $\htop_{1,2}$ and $\hat\htop_{1,2}$ in 
eqns. \req{hrs} and \req{hh0rs}, respectively.

\vskip 0.16in

{\it For} $\ket{\Delta, \htop}$
{\it chiral} we can set $\gamma=0$ and the h.w. 
conditions  on $\ket{\chi_{NS}}_1^{(0)ch}$ give the equations
               
\begin{eqnarray}
\alpha+\beta &=& 0\nonumber\\
\a{\ }\htop+\b{\ }{\ctop\over3} &=& 0\,.
\end{eqnarray}

\noi
Comparing these with eqns. \req{1nch} 
for $\Delta={\htop\over2}$, we see that the first and fourth
equations in \req{1nch} coincide now with the first one here, 
 while the third equation in \req{1nch}, which corresponds
 to the h.w. condition $~G_\half^+\ket\chi=0$, has disappeared,
the reason being that the complete equation reads
$~(\alpha-\beta-\gamma(2\Delta+\htop))G_{-\half}^+\ket{\Delta,\htop}=0$.
The solution to these equations is
${\ }\beta=-\alpha,{\ \ \ }\htop={\ctop\over 3}{\ }$, that is

\BE \ket{\chi_{NS}}_1^{(0)ch}= (L_{-1} - H_{-1})
{\ \ }\ket{\Delta=\ctop/6, {\ } \htop=\ctop/3}\,. \EE

\noi
Therefore, only the solution $(-\htop_{1,2})={\ctop\over3}$, denoted as
$(-\htop_{1,2}^{(0)})$ in eq. \req{h0rs},
remains after switching
on chirality on the primary state $\ket{\Delta,\htop}$, whereas the
singular vector $\ket{\chi_{NS}}_1^{(0)}$ vanishes for
$(-\hat\htop_{1,2})=-1$, as one can check in \req{eq1nch}.

\vskip 0.16in

{\it For} $\ket{\Delta, \htop}$
{\it antichiral} the h.w. 
conditions on $\ket{\chi_{NS}}_1^{(0)a}$ give the equations
               
\begin{eqnarray}
\alpha-\beta &=& 0\nonumber\\
\a{\ }\htop+\b{\ }{\ctop\over3} &=& 0\,.
\end{eqnarray}

\noi
Comparing these with eqns. \req{1nch} we see that now the 
last equation in \req{1nch}, which corresponds
 to the h.w. condition $~G_\half^-\ket\chi=0$, has disappeared,
the reason being that the complete equation reads
$~(\alpha+\beta+\gamma(2\Delta-\htop+2))
G_{-\half}^-\ket{\Delta,\htop}=0$.
The solution to these equations is
${\ }\beta=\alpha,{\ \ \ }\htop=-{\ctop\over 3}{\ }$, that is

\BE \ket{\chi_{NS}}_1^{(0)a}= (L_{-1} + H_{-1})
{\ \ }\ket{\Delta=\ctop/6, {\ } \htop=-\ctop/3}\,. \EE

\noi
Therefore, only the solution ${\ }\htop_{1,2}=-{\ctop\over3}{\ }$, denoted
as $\htop_{1,2}^{(0)}$ in eq. \req{h0rs},
remains after switching
on antichirality on the primary state $\ket{\Delta,\htop}$, whereas
the singular vector $\ket{\chi_{NS}}_1^{(0)}$ vanishes for
${\ }\hat\htop_{1,2}=1{\ }$, as one can check in \req{eq1na}.

Observe that $\ket{\chi_{NS}}_1^{(0)ch}$ and
$\ket{\chi_{NS}}_1^{(0)a}$ are symmetric under the interchange
$H_m \leftrightarrow -H_m, {\ } G^+_r \leftrightarrow G^-_r{\ }$,
as expected. This also happens for the singular vectors
$\ket{\chi_{NS}}_1^{(0)}$ specialized to the cases
 $\Delta = {\htop\over2}$ and $\Delta = -{\htop\over2}$, eqns. 
\req{eq1nch} and \req{eq1na} (one has to take into account that the
commutation $G^+_{-1/2} G^-_{-1/2} \rightarrow G^-_{-1/2} G^+_{-1/2}$
produces a global minus sign). As a matter of fact, the singular vectors
$\ket{\chi_{NS}}_1^{(0)ch}$ and $\ket{\chi_{NS}}_1^{(0)a}$ are nothing
but the singular vectors $\ket{\chi_{NS}}_1^{(0)}$
given by the solutions \req{eq1nch} and \req{eq1na} (for
$\htop=\pm\ctop/3$) after dropping the term
$G_{-1/2}^+G_{-1/2}^-$ due to the chirality and antichirality
conditions on $\ket{\D,\htop}$.

\vskip .20in
\noi
{\it Subsingular vector behaviour}

\vskip .08in
\noi
Now let us see that the charged subsingular vectors at level $3\over2$ 
behave properly; that is, they are not descendant null vectors
of the uncharged
singular vectors which sit at level 1 in the complete Verma modules.
As a matter of fact,
we will see that these charged subsingular vectors, which are 
null, can descend down to the uncharged
singular vectors, but not the other way around.

To see this in some detail
let us write the uncharged singular vectors at level 1
for $\htop=\pm 1 {\ }, \Delta= \mp {\htop\over2}=-{1\over2}{\ },$
given by eqns. \req{eq1na} and \req{eq1nch}. 
 For $\htop=1,{\ }\Delta=-{1\over2}{\ },$ the singular vector is 

\BE \ket{\chi_{NS}}_1^{(0)} = G^+_{-1/2} G^-_{-1/2}{\ \ }
    \ket{\Delta=-{\half},{\ }\htop=1}\,. \label{v1n1} \EE

\noi
Its level ${3\over2}$ secondary with relative charge $q=1$ vanishes
since ${\ }G^+_{-1/2}{\ }\ket{\chi_{NS}}_1^{(0)} =0{\ }$.
Similarly, for
$\htop=-1,{\ }\Delta=-{1\over2}{\ },$ the singular vector is 

\BE \ket{\chi_{NS}}_1^{(0)} = G^-_{-1/2} G^+_{-1/2}{\ \ }
    \ket{\Delta=-{\half},{\ }\htop=-1}\,, \label{v1n-1} \EE

\noi
and its level ${3\over2}$ descendant with relative charge $q=-1$ vanishes.
(It is also straightforward to see that the uncharged singular
vectors \req{v1n1} and \req{v1n-1} vanish when one imposes
antichirality and chirality on the primary states, respectively,
as we said before.)

On the other hand, the charged singular vectors for 
$\htop=\pm 1 {\ }, \Delta= \mp {\htop\over2}=-{1\over2}{\ },$  
given by \req{sol3a} and \req{sol3ch}, are

\BE 
\ket{\chi_{NS}}_{3\over 2}^{(1)a}=
({\ctop-3\over6}{\ } L_{-1}G_{-1/2}^+ -
 H_{-1}G_{-1/2}^+{\ }+{\ } G_{-3/2}^+)
{\ \ }\ket{\Delta=-\half,{\ } \htop=1}  \label{v3a1}
\EE
 
\noi
and

\BE 
\ket{\chi_{NS}}_{3\over 2}^{(-1)ch}=
({\ctop-3\over6}{\ } L_{-1}G_{-1/2}^- +
 H_{-1}G_{-1/2}^-{\ }+{\ } G_{-3/2}^-)
{\ \ }\ket{\Delta=-\half,{\ } \htop=-1}  \label{v3c-1}
\EE

If we now switch off antichirality (chirality) on the primary
state $\ket{\Delta,\htop}$, then the h.w. condition
$G^-_{1/2} \ket{\chi_{NS}}_{3\over 2}^{(1)a}=0{\ }$
($G^+_{1/2} \ket{\chi_{NS}}_{3\over 2}^{(-1)ch}=0{\ }$)
is not satisfied anymore, although the vectors are still null,
\ie\ have zero norm, as the reader can verify. However, these
vectors cannot be descendant states of the uncharged 
singular vectors \req{v1n1} and \req{v1n-1}, as we have just discussed,
nor are they descendant states of any level $\half$ singular vectors.

Interesting enough, these level $3\over2$ subsingular vectors do descend to the
level 1 uncharged singular vectors \req{v1n1} and \req{v1n-1}
under the action of $G^-_{1/2}{\ }$ and $G^+_{1/2}{\ }$,
 respectively, as is easy to check, but not the other
way around. The reason is that the singular
vectors \req{v1n1} and \req{v1n-1} do not build complete Verma
modules, as is the usual case for the singular vectors of
the N=2 superconformal algebra. 
 
\vskip .20in
\noi
{\it Singular vector ``transmutation"}

\vskip .08in
\noi
We have seen that the uncharged singular vectors $\ket{\chi_{NS}}_1^{(0)}$,
for $\htop=\pm\hat\htop_{1,2} =\pm1{\ }, \Delta = -\half$,
vanish when one switches on antichirality and chirality, respectively, on 
$\ket{\Delta, \htop}$, while the charged subsingular vectors become singular
vectors: $\ket{\chi_{NS}}_{3\over2}^{(1)a}$ for
$\htop= \hat\htop_{1,2}=1{\ }$
and $\ket{\chi_{NS}}_{3\over2}^{(-1)ch}$ for
$\htop=-\hat\htop_{1,2}=-1{\ }$. The other way around
when one switches off antichirality or chirality on
$\ket{\Delta, \htop}$; that is, the uncharged singular
vectors appear and the charged singular vectors ``dissappear" (they are
not singular vectors anymore), for the values of $\htop$ indicated before,
as if a mechanism of ``singular vector conservation" or
``transmutation" underlies the process.

To shed more light on the issue let us
analyze the behaviour of the uncharged singular vector
$\ket{\chi_{NS}}_1^{(0)}$  and its level $3\over2$ 
descendants near the limits $\htop \rightarrow \pm1, {\ }
\Delta \rightarrow - {\half}$.

Let us start with  $\htop$ near 1. Thus we set
${\ }\htop=1+\epsilon{\ },{\ }\Delta=-{1\over 2}(1+\delta){\ },$
$\Delta$ and $\htop$ satisfying the quadratic vanishing
surface relation, which results in

\BE {\ctop\over3}={\delta-2\epsilon-\epsilon^2\over\delta}\,. \label{ced} \EE

\noi
The vector $\ket{\chi_{NS}}^{(0)}_1$   is expressed now as

 \BE
\ket{\chi_{NS}}_1^{(0)}=(\epsilon L_{-1}+\delta H_{-1}+
G_{-1/2}^+G_{-1/2}^-)\ket{\Delta,\htop}\,.
\EE

\noi
Its charge $q=1$ descendant at level $3\over2$, which we denote as
$\ket\Ups^{(1)}_{3\over2}$, is a null vector which, in 
principle, is not h.w.

\BE 
\ket\Ups^{(1)}_{3\over2} = 
G_{-1/2}^+ \ket{\chi_{NS}}_1^{(0)} =
(\epsilon L_{-1}G_{-1/2}^++\delta H_{-1}G_{-1/2}^+
-\delta G_{-3/2}^+)\ket{\Delta,\htop}\,. \label{nfn}
\EE

\noi
Now comes a subtle point. In principle
we can normalize $\ket\Ups^{(1)}_{3\over2}$ in the same way as
$\ket{\chi_{NS}}^{(1)a}_{3\over2}$, \ie\ dividing all the
coefficients by $(-\delta)$ so that the coefficient
of $G^+_{-3/2}$ is 1, resulting in

\BE \ket\Ups^{(1)}_{3\over2} = 
((-{\epsilon\over\delta}) L_{-1}G_{-1/2}^+-H_{-1}G_{-1/2}^+
+G_{-3/2}^+)\ket{\Delta,\htop}\,.  \label{chied}
\EE
 
\noi
However, these two normalizations are not
equivalent when taking the limit $(\htop=1, \Delta=-\half){\ }$, \ie\ 
${\ }(\e \rightarrow 0, \delta \rightarrow 0){\ }$. Namely
 $\ket\Ups^{(1)}_{3\over2}$ vanishes with the first
normalization \req{nfn} whereas with the second normalization \req{chied}
it becomes the h.w. singular vector  $\ket{\chi_{NS}}^{(1)a}_{3\over2}{\ }$, 
since $(-{\epsilon\over\delta})$ turns into
$({\ctop-3\over6})$, as can be deduced easily from \req{ced}.                       

It seems that the two normalizations distinguish whether the
primary state $\ket{\Delta,\htop}$ approaches an antichiral or a
 non-chiral state as $\htop \rightarrow 1, \Delta 
\rightarrow -\half$. This is indeed true, the reason is that if we 
normalize $\ket{\chi_{NS}}^{(0)}_1$ in the same way as its descendant
$\ket\Ups^{(1)}_{3\over2}$, then in the
second normalization, \ie\ dividing by $(-\delta)$, it blows up
approaching the limit $(\e \rightarrow 0, \delta \rightarrow 0)$
unless the term $G^+_{-1/2} G^-_{-1/2}$ goes away, exactly what
happens if $\ket{\Delta,\htop}$ is antichiral. 
But the resulting uncharged vector without the term
$G^+_{-1/2} G^-_{-1/2}$  is not a
 singular vector anymore. The action
of $G^+_{-1/2}$ on this vector results
 precisely in the charged singular vector
 $\ket{\chi_{NS}}^{(1)a}_{3\over2}{\ }$. 

We see therefore that, in the limit  $(\htop=1, \Delta=-\half){\ }$,
 the first normalization produces the uncharged singular vector
at level 1 ${\ }\ket{\chi_{NS}}^{(0)}_1$, with a 
vanishing charge $q=1$ descendant
at level $3\over2$, whereas the second normalization produces the 
charge $q=1$ singular vector
at level $3\over2$ ${\ }\ket{\chi_{NS}}^{(1)a}_{3\over2}{\ }$.

Repeating this analysis for the case $\htop$ near $-1$ we find that
in the limit  $(\htop=-1, \Delta=-\half){\ }$, the first 
 normalization produces the uncharged singular vector at level 1 
${\ }\ket{\chi_{NS}}^{(0)}_1$, with a vanishing charge $q=-1$ descendant
at level $3\over2$, whereas the second normalization produces the 
charge $q=-1$ singular vector
at level $3\over2$ ${\ }\ket{\chi_{NS}}^{(-1)ch}_{3\over2}{\ }$.


\begin{thebibliography}{9}
\def\NPB{Nucl. Phys. B}
\def\PLB{Phys. Lett. B}
\def\MPLA{Mod. Phys. Lett. A}



\bibitem{Ade} M. Ademollo et al., Phys. Lett. B62 (1976) 105;
  \NPB111 (1976) 77; \NPB114 (1976) 297; \\
 L. Brink and J.H. Schwarz, \NPB121 (1977) 285; 

\bibitem{FrTs} E.S. Fradkin and A.A. Tseytlin, \PLB106 (1981) 63;
 \PLB162 (1985) 295; \\
 S.D. Mathur and S. Mukhi, Phys. Rev. D 36 (1987) 465;
\NPB302 (1988) 130; 

\bibitem{OoVa} H. Ooguri and C. Vafa, \NPB361 (1991) 469; 
 \NPB367 (1991) 83

\bibitem{Marcus} J. Gomis and H. Suzuki, \PLB278 (1992) 266; \\
N. Marcus, talk at String Theory Workshop, Rome (1992), hep-th/9211059  

\bibitem{BeSe2} B.~Gato-Rivera and A.~M.~Semikhatov, \PLB293 (1992) 72,
 Theor. Mat. Fiz. 95 (1993) 239, Theor. Math. Phys. 95 (1993) 536

\bibitem{BLNW} M. Bershadsky, W. Lerche, D. Nemeschansky and N.P. Warner,
 \NPB401 (1993) 304

\bibitem{BeSe3} B.~Gato-Rivera  and A.~M.~Semikhatov, \NPB408 (1993) 133

\bibitem{Sem} A.M. Semikhatov, Mod. Phys. Lett. A9 (1994) 1867

\bibitem{BJI2} B. Gato-Rivera and J.I. Rosado, Phys. Lett. B346 (1995) 63

\bibitem{BJI3} B. Gato-Rivera and J.I. Rosado, Phys. Lett. B369 (1996) 7

\bibitem{BJI6} B. Gato-Rivera and J.I. Rosado, ``Families of Singular and
Subsingular Vectors of the Topological N=2
Superconformal Algebra", hep-th/9701041 (1997)

\bibitem{BFK} W. Boucher, D. Friedan and A. Kent, Phys. Lett. B172 (1986) 316
     
\bibitem{Nam} S. Nam, \PLB172 (1986) 323

\bibitem{KaMa3} M. Kato and S. Matsuda, \PLB184 (1987) 184

\bibitem{EY} T.~Eguchi and S.~K.~Yang, \MPLA5 (1990) 1653

\bibitem{W-top} E.~Witten, Commun. Math. Phys. 118 (1988) 411;
 \NPB340 (1990) 281

\bibitem{DVV} R. Dijkgraaf, E. Verlinde and H. Verlinde, \NPB352
(1991) 59

\bibitem{PDiV} P. Di Vecchia et al., \PLB162 (1985) 327

\bibitem{SS} A. Schwimmer and N. Seiberg, \PLB184 (1987) 191

\bibitem{LVW} W.~Lerche, C.~Vafa and N.~P.~Warner,
 \NPB324 (1989) 427

\bibitem{Kir1} E.B. Kiritsis, Phys. Rev. D36 (1987) 3048 

\bibitem{BJI4} B. Gato-Rivera and J.I. Rosado, 
 Mod. Phys. Lett. A11 (1996) 423 

\bibitem{BJI7} B. Gato-Rivera, ``The Even and the Odd Spectral Flows 
on the N=2 Superconformal Algebras", hep-th/9707211 (1997) 

\bibitem{Kir2} E.B. Kiritsis, Int. J. Mod. Phys A3 (1988) 1871

\bibitem{YuZh} M. Yu and H.B. Zheng, \NPB288 (1987) 275

\bibitem{BJI8} B. Gato-Rivera and J.I. Rosado, ``Families of Singular and
Subsingular Vectors of the Antiperiodic and Periodic N=2
Superconformal Algebras", to appear in hep-th

\bibitem{KD} A. Kent and M. D\"orrzapf, private communication 

\bibitem{FF} B.L. Feigin and D.B. Fuchs, ``Representations of the 
Virasoro algebra" in ``Representations of Lie groups and related Topics",
Gordon and Breach, 1990


\end{thebibliography}
\end{document}